\documentclass{aa}
\usepackage{graphicx}

\def\ltsim{\raise 2pt \hbox {$<$} \kern-1.1em \lower 4pt \hbox {$\sim$}}
\def\ltapprox{\raise 2pt \hbox {$<$} \kern-1.1em \lower 5pt \hbox {$\approx
$}}
\def\gtsim{\raise 2pt \hbox {$>$} \kern-1.1em \lower 4pt \hbox {$\sim$}}
\def\gtapprox{\raise 2pt \hbox {$>$} \kern-1.1em \lower 5pt \hbox {$\approx
$}}

\begin{document}

\title{GMRT Radio Halo Survey in galaxy clusters at z = 0.2 -- 0.4.}
\subtitle{II.The eBCS clusters and analysis of the complete sample} 
\author{T.~Venturi\inst{1} \and
S.~Giacintucci\inst{1,}\inst{2,}\and 
D.~Dallacasa\inst{3,}\inst{1}\and 
R.~Cassano\inst{3,}\inst{1}\and
G.~Brunetti\inst{1}\and
S.~Bardelli\inst{4} \and
G.~Setti\inst{3,}\inst{1}
}

\institute
{
INAF -- Istituto di Radioastronomia, via Gobetti 101, I-40129, Bologna, Italy 
\and
Harvard--Smithsonian Center for Astrophysics, 60 Garden Street, Cambridge, 
MA02138, USA 
\and
Dipartimento di Astronomia, Universit\`a di Bologna,
via Ranzani 1, I--40126, Bologna, Italy
\and
INAF -- Osservatorio Astronomico di Bologna,
via Ranzani 1, I--40127 Bologna, Italy
}
\date{}
%
%
\abstract
{}
{We present the results of the GMRT cluster radio halo survey. 
The main purposes of our observational project are to measure which 
fraction of massive galaxy  clusters in the redshift range z=0.2--0.4 
hosts a radio halo, and to constrain the expectations of the particle 
re--acceleration model for the origin of the non--thermal radio emission.} 
{We selected a complete sample of 50 clusters in the X--ray band from the 
REFLEX (27) and the eBCS (23) catalogues. In this paper we present Giant 
Metrewave Radio Telescope (GMRT) observations at 610 MHz for all clusters 
still lacking high sensitivity radio information, i.e. 16 eBCS 
and 7 REFLEX clusters, thus completing the radio information for the whole 
sample. The typical sensitivity in our images is in the range 
1$\sigma \sim 35-100~\mu$Jy b$^{-1}$.}
{We found a radio halo in A\,697, a diffuse peripheral source of unclear
nature in A\,781, a core--halo source in Z\,7160, a candidate radio halo
in A\,1682 and ``suspect'' central emission  in Z\,2661.
Including the literature information, a total of 10 clusters in the
sample host a radio halo. A very important result of our work is that 25 
out of the 34 clusters observed with the GMRT do not host extended central 
emission at the sensitivity level of our observations, and for 20 of them 
firm upper limits to the radio power of a giant radio halo were derived.
The GMRT Radio Halo Survey shows that radio halos are not common,   
and our findings on the fraction of giant radio halos
in massive clusters are consistent with the statistical expectations 
based on the re--acceleration model. Our results favour primary to 
secondary electron models.}
{}

\keywords{Radio continuum: galaxies -- galaxies: clusters: general -- galaxies:
clusters: individual: }

\maketitle
\section{Introduction and the GMRT radio halo survey}\label{sec:intro}

Radio and X--ray observations of galaxy clusters prove that 
the intracluster medium (ICM) in clusters of galaxies is a mixture
of hot plasma, magnetic field and relativistic particles.
While X--ray observations reveal the presence of diffuse hot gas, 
the spectacular synchrotron radio emission extended on the Mpc scale and 
observed in a growing number of massive clusters is the signature 
of the the presence of relativistic electrons (with Lorentz factor 
$\gamma >>$ 1000) and magnetic fields spread over the whole
cluster volume (e.g. Feretti \cite{feretti05a}). Recent studies revealed
that the magnetic fields in galaxy clusters are weak, with strength in the 
range $\sim 0.1 - 1~\mu$G (for recent reviews see Govoni \& Feretti 
\cite{gf04} and Ferrari et al. \cite{ferrari08}).
The recently discovered hard X--ray tails in excess of the thermal 
Bremsstrahlung spectrum in a few galaxy clusters (Fusco--Femiano et al. 
\cite{fusco04}; Rephaeli et al. \cite{rephaeli08}; Fusco--Femiano \& Orlandini
\cite{fusco08}) are also considered a piece of evidence in favour of a 
non--thermal component in the ICM (see also Rossetti \& Molendi
\cite{rm04} for further discussion on this issue).

The extended cluster radio emission may take the form of {\it radio halos},
{\it relics} and {\it core--halos} (or mini--halos). While the latter 
reach extensions of the order of \ltsim 500 kpc, are associated with  
the dominant galaxy in cooling core clusters and are thought to be related to 
the radio emission of the central AGN, halos and relics are much larger
in size (reaching and exceeding the Mpc scale) and are not associated with AGN 
activity in individual galaxies.
{\it Radio halos} are usually located at the centre of galaxy clusters, 
and show a fairly regular morphology; {\it relics} are found at the 
cluster periphery, are highly polarized and exhibit a variety of radio 
morphologies, the most common being sheet, arc, toroids.
There is some consensus on the fact that the origin of radio relics resides 
in cluster mergers and/or matter accretion: the strong peripheral shocks 
developing during these energetic events may be efficient particle 
accelerators (Sarazin \cite{sarazin99}; Ryu \cite{ryu03}; 
Pfrommer \cite{pfrommer06}). 

The origin of the synchrotron radio emission in halos is 
still an open problem, since the life--time of the electrons emitting
synchrotron radiation is much shorter than the diffusion time necessary
to cover their large extent, typically of the order of the Mpc.
\\
Historically, two main possibilities have been suggested to account for
the existence of radio halos: (1) ``re--acceleration models'', or
``primary electron'' models, where 
electrons are re--accelerated in--situ, due to the turbulence injected in the
cluster volume by massive  merger events (e.g. Brunetti et al. 
\cite{brunetti01}; Petrosian \cite{petrosian01}; and the review papers by
Brunetti \cite{brunetti03}; Sarazin \cite{sarazin04}; Petrosian \& Bykov
\cite{pb08}); (2) ``secondary electron models'',
which predict that relativistic electrons are secondary products of
hadronic collisions between the cluster ICM and cosmic rays (e.g. Dennison
\cite{dennison80}; Blasi \& Colafrancesco \cite{bc99}; Dolag \& En\ss{}lin 
\cite{de2000}).

The re--acceleration scenario implies a tight connection between the 
process 
of hierarchical cluster formation and the non--thermal phenomena, and indeed  
there is observational evidence that radio halos are found in clusters with 
signatures of ongoing merging
processes (e.g. Buote \cite{buote01}; Schuecker et al. \cite{schuecker01}), 
and that the occurrence of radio 
halos increases with the X--ray luminosity of the parent clusters 
(Giovannini et al. \cite{giovannini99}; Cassano et al. \cite{cassano08}).
\\
Growing attention is given to the 
statistical properties of radio halos in the framework of this model.
Calculations 
were developed 
in order to model the connection between radio halos and the merging history
of galaxy clusters. Expectations were made for the occurrence of 
{\it giant}\footnote{Linear size  $\ge$ 1 Mpc as defined in CB05, with 
H$_0$=50 ~km~s$^{-1}$~Mpc$^{-1}$. 
This size corresponds to \gtsim 700 kpc with the cosmology assumed in this 
paper.} 
radio halos as a function of redshift and mass (Cassano \& Brunetti 
\cite{cb05}, hereinafter CB05; Cassano, Brunetti \& Setti \cite{cbs06}, 
hereinafter CBS06; Cassano et al. \cite{cassano08}). 
The predictions 
show that the bulk of giant radio halos is expected in the redshift 
range z=0.2--0.4, where a fraction of $\sim$35\% of massive clusters 
(M$~>~10^{15}$ M$_{\odot}$) may host such cluster radio sources.
\\
\\
In order to investigate the connection between cluster mergers and the 
presence of diffuse sources in galaxy clusters, and to test the 
statistical predictions, we selected a complete sample of 50 
massive galaxy 
clusters in the redshift range z = 0.2 -- 0.4, and carried out a deep pointed
radio survey with the Giant Metrewave Radio Telescope (GMRT, Pune, India) 
at 610 MHz, imaging all clusters in the sample lacking high sensitivity 
information. We refer to this project as the ``GMRT radio halo survey''.
In Venturi et al. (\cite{venturi07}, hereinafter Paper I) 
we reported the results on 11 clusters observed. 
\\
\\
In this paper we present the completion of the GMRT radio halo survey. 
The paper is organised as follows: in Section 2 we present the sample; in 
Section 3 we describe the radio observations; in Section 4 and 5 we present 
the results; in Section 6 we discuss the results of the 
GMRT radio halo survey as a whole; summary and conclusions in the light
of the re--acceleration model are given in Section 7. 
\\
The cosmology adopted in this paper is $H_0 =70$~km~s$^{-1}$~Mpc$^{-1}$, 
$\Omega_m$=0.3 and $\Omega_{\Lambda}$=0.7.
We assume S$\propto~\nu^{-\alpha}$ throughout the paper.

\section{The cluster sample}\label{sec:sample}

\begin{table*}[t]
\label{tab:sample1}
\caption[]{Cluster sample from the REFLEX and eBCS catalogues.}
\begin{center}
\begin{tabular}{lrrcrc}
\hline\noalign{\smallskip}
Cluster name   & RA$_{J2000}$ &  DEC$_{J2000}$ & z &  L$_{\rm X}$(0.1--2.4 keV) & Notes \\ 
               &              &                &   & $10^{44}$ erg s$^{-1}$     & \\
\noalign{\smallskip}
\hline\noalign{\smallskip}
REFLEX  Sample    &&&&& \\
\noalign{\smallskip}
\hline\noalign{\smallskip}

A\,2697 &  00 03 11.8 &  --06 05 10 & 0.2320 &   6.88 & Paper I \\
A\,2744 &  00 14 18.8 &  --30 23 00 & 0.3066 &  12.92 & (1) \\
A\,2813 &  00 43 24.4 &  --20 37 17 & 0.2924 &   7.62 & This paper \\
A\,141  &  01 05 34.8 &  --24 39 17 & 0.2300 &   5.76 & Paper I \\ 
A\,2895 &  01 18 11.1 &  --26 58 23 & 0.2275 &   5.56 & This paper \\ 
A\,209  &  01 31 53.0 &  --13 36 34 & 0.2060 &   6.29 & Paper I \\
A\,3088 &  03 07 04.1 &  --28 40 14 & 0.2537 &   6.95 & Paper I \\
RXCJ\,0437.1$+$0043   &  04 37 10.1 &   +00 43 38 & 0.2842 &   8.99 
& (2) \\
A\,521  &  04 54 09.1 &  --10 14 19 & 0.2475 &   8.18 & Paper I,(3) \\
RXCJ\,0510.7$-$0801   &  05 10 44.7 &  --08 01 06 & 0.2195 &   8.55 
& $^{\surd}$ \\
A\,3444 &  10 23 50.8 &  --27 15 31 & 0.2542 &  13.76 & Paper I  \\
RXCJ\,1115.8$+$0129   &  11 15 54.0 &   +01 29 44 & 0.3499 &  13.58 
& This paper \\
A\,1300 &  11 31 56.3 &  --19 55 37 & 0.3075 &  13.97 & (4) \\
RXCJ\,1212.3$-$1816   &  12 12 18.9 &  --18 16 43 & 0.2690 &   6.20 
& $^{\surd}$ \\
RXCJ\,1314.4$-$2515   &  13 14 28.0 &  --25 15 41 & 0.2439 &  10.94 
& Paper I,(2) \\
S\,780  &  14 59 29.3 &  --18 11 13 & 0.2357 &  15.53  & Paper I \\
RXCJ\,1504.1$-$0248   &  15 04 07.7 &  --02 48 18 & 0.2153 &  28.08 
& $^{\surd}$ \\
RXCJ\,1512.2$-$2254   &  15 12 12.6 &  --22 54 59 & 0.3152 &  10.19 
& This paper \\
RXCJ\,1514.9$-$1523   &  15 14 58.0 &  --15 23 10 & 0.2226 &   7.16 
& $^{\surd}$ \\
A\,2163 &  16 15 46.9 &  --06 08 45 & 0.2030 &  23.17 & (5),(6) \\
RXCJ\,2003.5$-$2323   &  20 03 30.4 &  --23 23 05 & 0.3171 &   9.25 
& Paper I \\
RXCJ\,2211.7$-$0350   &  22 11 43.4 &  --03 50 07 & 0.2700 &   7.42 
& $^{\surd}$ \\
A\,2485 &  22 48 32.9 &  --16 06 23 & 0.2472 &   5.10 & This paper \\ 
A\,2537 &  23 08 23.2 &  --02 11 31 & 0.2966 &  10.17 & Paper I \\
A\,2631 &  23 37 40.6 &  +00 16 36 & 0.2779  &   7.57 & Paper I \\
A\,2645 &  23 41 16.8 &  --09 01 39 & 0.2510 &   5.79 & This paper \\ 
A\,2667 &  23 51 40.7 &  --26 05 01 & 0.2264 &  13.65 & This paper \\
\noalign{\smallskip}
\hline\noalign{\smallskip}
eBCS Sample       &&&&& \\
\noalign{\smallskip}
\hline
\\
RXJ\,0027.6$+$2616 & 00 27 49.8 &  +26 16 26 & 0.3649 & 12.29  & This paper \\
A\,611   & 08 00 58.1 &  +36 04 41 & 0.2880 &  8.86  & This paper \\
A\,697   & 08 42 53.3 &  +36 20 12 & 0.2820 & 10.57 & This paper \\
Z\,2089  & 09 00 45.9 &  +20 55 13 & 0.2347 &  6.79 & This paper \\
A\,773   & 09 17 59.4 &  +51 42 23 & 0.2170 &  8.10 & (1) \\
A\,781   & 09 20 23.2 &  +30 26 15 & 0.2984 & 11.29 & This paper \\
Z\,2661  & 09 49 57.0 &  +17 08 58 & 0.3825 & 17.79 & This paper \\
Z\,2701  & 09 52 55.3 &  +51 52 52 & 0.2140 &  6.59 & This paper  \\
A\,963   & 10 17 09.6 &  +39 01 00 & 0.2060 &  6.39 & This paper \\
A\,1423  & 11 57 22.5 &  +33 39 18 & 0.2130 &  6.19 & This paper \\
Z\,5699  & 13 06 00.4 &  +26 30 58 & 0.3063 &  8.96 & This paper \\
A\,1682  & 13 06 49.7 &  +46 32 59 & 0.2260 &  7.02 & This paper \\
Z\,5768  & 13 11 31.5 &  +22 00 05 & 0.2660 &  7.47 & This paper \\
A\,1758a & 13 32 32.1 &  +50 30 37 & 0.2800 & 12.26 & (7)  \\
A\,1763  & 13 35 17.2 &  +40 59 58 & 0.2279 &  9.32 & VLA Archive   \\
Z\,7160  & 14 57 15.2 &  +22 20 30 & 0.2578 &  8.41 & This paper \\
Z\,7215  & 15 01 23.2 &  +42 21 06 & 0.2897 &  7.34 & This paper \\
RXJ\,1532.9$+$3021 & 15 32 54.2 &  +30 21 11 & 0.3450 & 16.49 & This paper \\
A\,2111  & 15 39 38.3 &  +34 24 21 & 0.2290 &  6.83  & VLA Archive  \\
A\,2219  & 16 40 21.1 &  +46 41 16 & 0.2281 & 12.73  & (8) \\
A\,2261  & 17 22 28.3 &  +32 09 13 & 0.2240 & 11.31  & VLA Archive \\
A\,2390  & 21 53 34.6 &  +17 40 11 & 0.2329 & 13.43  & (8) \\
RXJ\,2228.6$+$2037 & 22 28 34.4 &  +20 36 47 & 0.4177 & 19.44 & This paper  \\
\noalign{\smallskip}
\hline\noalign{\smallskip}
\end{tabular}
\end{center}
References: Paper I: Venturi et al. \cite{venturi07};
(1) Govoni et al. \cite{govoni01}; (2) Feretti et al. 
\cite{feretti05}; (3) Giacintucci et al. \cite{g06}; (4) A\,1300 Reid et al. 
\cite{reid99}; (5) Herbig \& Birkinshaw \cite{herbig94}; 
(6) Feretti et al. \cite{feretti01}; (7) Giovannini et al. \cite{giovannini06};
(8) Bacchi et al. \cite{bacchi03};
$^{\surd}$ clusters which are part of the GMRT cluster Key Project 
(P.I. Kulkarni).

\end{table*}
%
%

From the ROSAT--ESO Flux Limited X--ray galaxy cluster catalogue (REFLEX, 
B\"ohringer et al. \cite{boeringer04}) and from the extended ROSAT
Brightest Cluster Sample catalogue (BCS, Ebeling et al. \cite{ebeling98},
\cite{ebeling00}) we selected all clusters satisfying the following 
constraints:

\begin{itemize} 
\item[1)] L$_{\rm X}$(0.1--2.4 keV) $>$ 5 $\times$ 10$^{44}$ erg s$^{-1}$;

 
\item[2)] 0.2 $<$ z $\le$ 0.4;

\item[3)] $-30^{\circ} < \delta < +2.5^{\circ}$ for the REFLEX
sample; $15^{\circ} < \delta < 60^{\circ}$ for the eBCS sample.

\end{itemize} 

The limit in X--ray luminosity comes from the need to 
select massive clusters, which are expected to host a giant radio halo
(Cassano, Brunetti \& Setti \cite{cassano04} and CB05).
The limit in L$_{\rm X}$ corresponds to a limit in virial mass 
M$_{\rm V} > 1.4 \times 10^{15}$ M$_{\odot}$, assuming the relation
L$_{\rm X}$ -- M$_{\rm V}$ derived in CBS06 (see also Paper I).
\\
The value $\delta = 2.5^{\circ}$ is the REFLEX declination limit;
the lower limit $\delta > -30^{\circ}$ was chosen in order to ensure 
good u--v coverage with the GMRT.  
In order to reach a compromise between the need to obtain a large
sample and the observational effort necessary to complete the requested
radio information, from the eBCS catalogue we selected all clusters in
the declination range  $15^{\circ} < \delta < 60^{\circ}$. 
The final full sample includes 50 clusters, 27 from the REFLEX catalogue 
and 23 from the eBCS. The complete list of clusters is reported in Table 1, 
where we provide the following
information: (1) cluster name (either from optical or from X--ray catalogues); 
(2) RA$_{\rm J2000}$ and (3) DEC$_{\rm J2000}$; (4) redshift; (5) X--ray 
luminosity; (6) notes on the radio information.
The upper part of the Table refers to the REFLEX clusters (see also Paper I), 
the lower part to the eBCS catalogue. Note that the cluster RXCJ\,2228.6+2037
has z=0.4177, i.e. just above our selection limit. We included it in
the sample observed at the GMRT, but it has not been considered in
our discussion (Section 6) and in the statistical analysis carried out in
Cassano et al . \cite{cassano08}.

\section{Radio Observations and data reduction}\label{sec:obs}

We carried out GMRT observations at 610 MHz for the 23 clusters listed in
Table 1 still lacking proper imaging at low resolution to assess the presence
of extended radio emission on the cluster scale, i.e. 7 REFLEX and 16 eBCS 
clusters (see last column in the Table).  
The observations were performed during three observing runs.
Most of the clusters (21) were observed during the period between 
30 September 2005 and 04 October 2005; A\,2645 and A\,2667 were observed 
on 27 August 2006;
A\,1682 was observed during the 2005 run, but it was reobserved on 05 
December 2006, 
due to the poor quality of the first dataset, which did not allow proper 
imaging.
\\
Each cluster was observed for a total of 2.5 -- 3 hours, with a duty cycle 
of 24 minutes on the target and 6 minutes on the phase calibrator. 
Depending on the allocated LST intervals, the observations for each cluster
covered a hour angle in the range 3 -- 5 hours.
\\
The data were recorded using simultaneously the upper and lower side bands 
(USB and LSB respectively), for a total bandwidth of 32 MHz. The default 
spectral line observing was performed, with each individual band divided into 
128 channels, with spectral resolution of 125 kHz/channel.
\\
The data analysis was carried out using the standard tasks of the 
NRAO AIPS (Astronomical 
Image Processing System) package. After bandpass calibration and removal of 
the channels at the edges of the band, the remaining 94 channels were averaged 
into 6 channels. The USB and LSB datasets were calibrated and self--calibrated 
independently, and the final images were obtained on the image plane by 
combining 
the final images from each individual dataset. Multifield imaging was carried 
out in each step of the data reduction, to account for the non--coplanarity
of the sky within the large field of view of the primary beam at 610 MHz.
\\
The full resolution of the GMRT at 610 MHz is $\sim 5^{\prime\prime}$ and the
largest nominal detectable structure is 17$^{\prime}$. This scale is much 
larger than the typical angular scale of Mpc--size radio halos in the redshift
interval under consideration here, which ranges from $\sim 3^{\prime}$ at 
z=0.4 to $\sim 5^{\prime}$ at z=0.2. This ensures the detection of the
Mpc size radio sources we are investigating (see Section 4 for a detailed
discussion on this point).
\\
Beyond the full resolution images, produced with uniform weighting and
no tapering, for each field we produced images with different resolutions, 
tapering the u--v data by means of the parameters {\tt robust} and 
{\tt uvtaper} in the task IMAGR. The typical ``low'' resolution
is of the order of 20$^{\prime\prime}~-~25^{\prime\prime}$.
\\
The rms noise (1$\sigma$ level) in the final images ranges 
from $\sim$ 40 $\mu$Jy b$^{-1}$ in the best cases, to 100 -- 140 
$\mu$Jy b$^{-1}$ in the worst. Most of the fields have an rms of the order of 
50 -- 80 $\mu$Jy b$^{-1}$.
For each cluster the rms in the full resolution and tapered 
images are comparable.  The quality of the final images depends on the 
usable bandwidth (in a number of cases only one of the two bands provided good 
data) and on the presence of strong sources in the proximity of the field 
centre. We estimate that the residual amplitude calibration uncertainty is 
of the order of \ltsim 5\%.
\\
In Table \ref{tab:obs} we list the convolution beam of the final images 
produced at the lowest resolution (column 2) and referred to in this paper
(Section 4), the average rms, estimated over regions far from strong 
sources (column 3); a note concerning the detection of cluster scale radio 
emission (column 4).
%
%
%
\begin{table} 
\caption[]{GMRT observations.}
\label{tab:radiodata}
\begin{center}
\begin{tabular}{llcc}
\hline\noalign{\smallskip}
Cluster  & Beam, PA  & rms    &  Notes$^{\diamond}$ \\ 
         & arcsec, $^{\circ}$ & $\mu$Jy b$^{-1}$ &  \\
\noalign{\smallskip}
\hline\noalign{\smallskip}
RXJ\,0027.6$+$2616  & $15.4\times11.2$, $~$53.6 &  65 & \\
A\,2813             & $13.9\times11.6$, $~$29.6 & 130 & \\
A\,2895$^{\star}$   & $~9.0\times~5.1$,$~~$36.3 & 140 & \\
A\,611              & $13.5\times12.0$, $~$60.0 &  50 & \\
A\,697              & $40.0\times35.0$, $~$37.0 &  50 & GRH \\
Z\,2089             & $15.4\times13.7$, $~$14.0 &  45 & \\
A\,781              & $18.0\times15.0$, $~~$0.0 &  50 & DE \\
Z\,2701             & $24.5\times16.3$, $~$43.0 &  75 & \\
Z\,2661             & $20.0\times20.0$, $~~$0.0 &  65 & C \\
RXCJ\,1115.8$+$0129 & $17.0\times10.0$, $~~$0.0 &  45 & \\
A\,963$^{\star}$    & $~6.0\times~4.5$,$~~$65.0 & 150 & \\
A\,1423             & $22.5\times14.0$, $~$45.5 &  85 & \\
Z\,5699             & $20.0\times13.6$, $~$57.2 &  75 & \\
A\,1682             & $20.3\times13.9$, $~$86.0 &  80 & C \\
Z\,5768             & $20.6\times16.8$, $~$39.1 &  70 & \\
Z\,7160             & $17.0\times14.0$, $~~$0.0 &  40 & c--H \\
Z\,7215             & $23.0\times18.0$, $-$0.9  & 100 & \\
RXCJ\,1512.2$-$2254 & $19.6\times16.9$, $~$23.6 &  80 & \\
RXJ\,1532.9$+$3021  & $24.0\times21.0$, $~$19.9 &  80 & \\
RXJ\,2228.6$+$2037  & $22.5\times18.0$, $-$5.8  &  80 & \\
A\,2485             & $16.1\times12.7$, $~$35.3 & 130 & \\
A\,2645$^{\star}$   & $10.8\times~6.4$, $~$32.5 & 250 & \\
A\,2667             & $17.0\times13.7$, $~$33.1 & 100 & \\
\hline{\smallskip}
\end{tabular}
\end{center}
$^{\star}$ Due to the presence of strong confusing sources  low
resolution images could not be produced.
\\
$^{\diamond}$ Notes on the radio emission: GRH=giant radio halo; 
DE=diffuse emission at the cluster periphery; C=candidate emission
at the cluster centre (positive residuals); c--H=core--halo. 
\label{tab:obs}
\end{table}
%
%
%
%
%
%
%
\begin{table}[t]
\label{tab:limits}
\caption[]{Radio power upper limits for all undetected radio halos in the 
GMRT Radio Halo Survey}
\begin{center}
\begin{tabular}{lcrl}
\hline\noalign{\smallskip}
Cluster Name   & z &  rms & Log$~$P$_{\rm 610~MHz}$  \\ 
               &   & $\mu$Jy b$^{-1}$  & W Hz$^{-1}$ \\
\noalign{\smallskip}
\hline\noalign{\smallskip}
A\,2697 &  0.2320 &  80  & 24.05 \\ 
A\,141  &  0.2300 &  90  & 24.07 \\
A\,3088 &  0.2537 &  65  & 24.07 \\ 
RXCJ\,1115.8$+$0129 &  0.3499 &  45  & 24.10$^{\star}$\\
S\,780  &  0.2357 &  65  & 24.02 \\
RXCJ\,1512.2$-$2254 &  0.3152 &  80  & 24.20\\
A\,2537 &  0.2966 &  65  & 24.15$^{\star}$ \\
A\,2631 &  0.2779 &  60  & 24.05 \\
A\,2667 &  0.2264 & 100  & 24.09 \\
\noalign{\smallskip}
\hline
\\
RXCJ\,0027.6$+$2616 &  0.3649 &  65  & 24.30 \\
A\,611   &  0.2880 &  50  & 24.07 \\
A\,781   &  0.2984 &  50  & 24.00$^{\star}$ \\
Z\,2089  &  0.2347 &  45  & 23.86 \\
Z\,2701  &  0.2140 &  75  & 23.98$^{\star}$ \\
A\,1423  &  0.2130 &  85  & 24.02 \\
Z\,5699  &  0.3063 &  75  & 24.20
\\
Z\,5768  &  0.2660 &  70  & 24.00 \\
Z\,7215  &  0.2897 & 100  & 24.20 \\
RXCJ\,1532.9$+$3021 &  0.3450 & 80  & 24.25 \\
RXCJ\,2228.6$+$2037 &  0.4177 & 80  & 24.40$^{\star}$ \\
\noalign{\smallskip}
\hline\noalign{\smallskip}
\end{tabular}
\end{center}
$^{\star}$ These upper limits are about $\sim$20\% deeper than those used 
in BVD07, due to improved images prepared for this paper.\\
\end{table}
%

\section{Undetections and upper limits}\label{sec:uplim}

Most of the clusters observed with the GMRT radio halo survey do not show 
any  hint of radio emission on the cluster scale at the sensitivity of the 
observations, as clear from Table \ref{tab:radiodata} and Paper I.
In order to provide quantitative observational information 
it is crucial to place firm upper limits 
to the radio power for the non detection of radio halos in our sample. 
From an observational point of view this means to place upper limits to the 
flux density from the final images. 

\subsection{The procedure}

We approached the problem following the steps described below.

\begin{itemize} 
\item[1)] We modelled the brightness profiles of well studied 
radio halos with sets of optically thin concentric spheres with different 
radius and flux density, and obtained ``families'' of radio halos with steps 
in total flux density and angular scale (see also Brunetti et al. 
\cite{brunetti07}, hereinafter BVD07, and Figure 1 therein). 
In particular,in each model the largest sphere contains $\sim$ 50\% of the 
total flux; 

\item[2)] each model for the brightness distribution, which we will refer to
as ``fake'' radio halo, was injected 
in a number of selected u--v datasets by means of the task UVMOD in AIPS. 
This task transforms the brightness distribution of the model into u--v 
components, adds them to the original data and produces a new ``modified'' 
u--v dataset;

\item[3)] imaging was carried out on each of the ``modified'' u--v datasets 
by means of the task IMAGR in AIPS. From the ``modified'' datasets 
we subtracted the individual sources in the central area of the cluster 
and imaged the residual emission;

\item[4)] we finally integrated the flux density (task TVSTAT) over an area 
as large as the angular scale of the injected radio halo.

\end{itemize}

Among the datasets presented here, for our analysis we chose  
representative clusters exploring the rms range of our observations
(see Table \ref{tab:obs}), whose final images do not show hints
of extended emission in the cluster central area. As a further test
of our procedure we injected the ``fake'' radio halos both at the cluster 
centre and in a nearby empty region (free from point sources and artifacts).
The largest angular scale
(LAS) of the ``fake'' radio halos explores the range 
from 180$^{\prime\prime}$ to 320$^{\prime\prime}$, so as to cover the angular 
scale corresponding to a linear size of 1 Mpc in the redshift interval 
0.2--0.4,
(see Section \ref{sec:obs}). We injected total flux density values 
S$_{\rm inj}$ from 3 to 30 mJy and  produced images at different 
resolutions, in the range $10^{\prime\prime} - 25^{\prime\prime}$.

\subsection{The results}

\begin{figure*}
\centering
\includegraphics[angle=0, width=19cm]{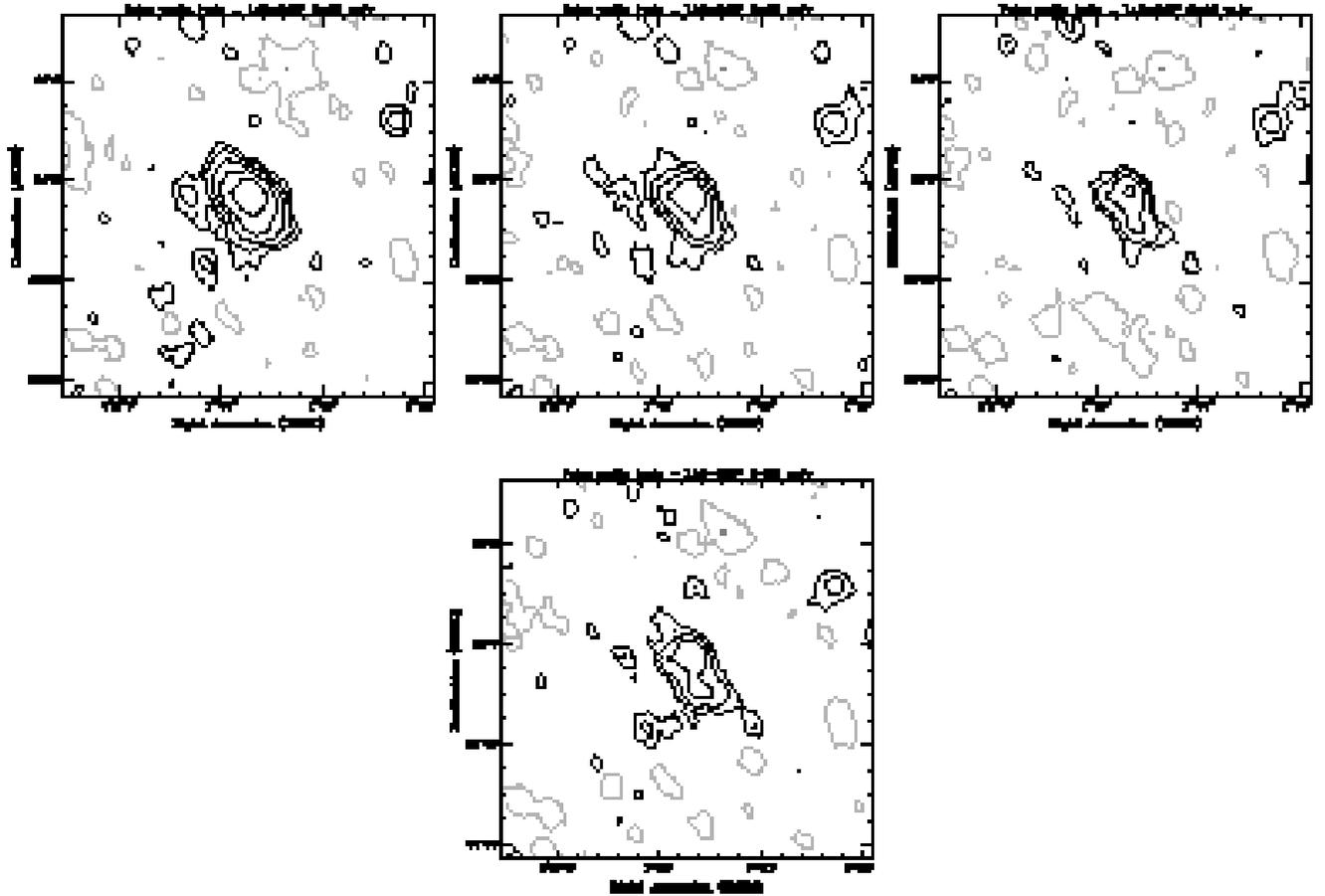}
\caption{610 MHz contours of fake radio halos. 
{\it Upper frame} -- Sequence 
with fixed LAS=240$^{\prime\prime}$ and decreasing flux density.
S$_{\rm inj}$ = 30, 20 and 15 going from left to right.
The restoring beam is $30^{\prime\prime}\times24^{\prime\prime}$;
the contour levels start from 0.15 mJy b$^{-1}$ and increase
with factor of 2. {\it Lower frame} -- An example with S$_{\rm inj}$= 20 
mJy LAS=320$^{\prime\prime}$, to be compared to the central 
panel in the upper frame. Restoring beam and contour levels
as in the upper frame.}
\label{fig:limits}
\end{figure*}

We found that the component of the ``fake'' radio halos with the largest 
angular size (i.e. the one with lowest brightness) 
is at least partly lost in the imaging process, the effect becoming more
important for faint flux densities and/or large angular sizes.
On the other hand, the central regions, i.e.
those with higher surface brightness, are usually well imaged. To
summarize, about up to $\sim$50\% of the injected flux density is lost.
As a consequence of this result,
the LAS in the imaged fake halos is usually smaller than the one 
injected. For clear detections, the measured LAS is of the order of 
$\sim 70$\% of the injected one.
\\
A conservative conclusion of our procedure is that above a value 
S$_{\rm inj}~\sim$ 12 mJy the presence of extended emission may be safely 
established over all ranges of angular scales considered in our analysis.
Injected flux densities of the order of 7--10 mJy result into positive
residuals with integrated flux density above the noise level
(i.e. $\sim 3-5\sigma$), which would lead to the conclusion of ``suspect''
emission. Therefore these values may be considered upper limits to
the radio halo flux density for those clusters where no evidence 
of central residual emission is found.
\\
As a visual example, in Figure \ref{fig:limits} we report a sequence 
of images with constant LAS and varying flux density (upper panels) and 
constant flux density and varying LAS (middle panel in the upper frame
and lower panel).
The panels in the figure highlight the dependence of the imaged ``fake'' halos 
on S$_{\rm inj}$ and LAS, and also show that the angular scales are 
smaller than those injected. 
\\  
It is important to point out that this procedure makes use of true
GMRT u--v datasets, therefore our conclusions on the detection limit of
radio halos apply to the observations presented here and in Paper I.
\\
\\
For each cluster observed with the GMRT (see  Table \ref{tab:sample1}) 
lacking extended emission in our images, we derived the upper limit to the 
radio power (expressed in W Hz$^{-1}$) following BVD07.
Our results are reported in Table 3, which provides upper
limits for 20 clusters. The upper part of the table refers to the REFLEX 
sample, the lower part to the eBCS. For a consistent analysis, the rms values
given for those clusters presented in Paper I refer to low resolution
images, and this accounts for some minor differences between the values
presented here and those in Table 2 of Paper I.
\\
Table 3 does not include the clusters with rms$\ge$130 $\mu$Jy b$^{-1}$, 
i.e. A\,2813, A\,2895, A\,2845, A\,2645, A\,963 (see Table 2). For all of
them only part of the visibilities could be used in the imaging process,
due to a number of different problems occurring during the observations,
and any limit set for them would not be significant. 
These five clusters were not included in the analysis carried
out in BVD07.
\\
We remind here that a radio halo was found in the clusters A\,209, 
RXCJ\,1314.4--2515 and RXCJ\,2003.5--2323; extended emission was found 
around the central galaxy in A\,3444 (Paper I); ``suspect'' extended emission
has been found in A\,521 at 327 MHz (Giacintucci et al. \cite{g08}) and
follow--up analysis is in progress (Brunetti et al. in prep.). 
Hence, these clusters are not included in the analysis.
From the list we also excluded A\,697, Z\,7160, A\,1682 and 
Z\,2661 (see Table \ref{tab:obs}), which will be presented in detail
in the next Section. 
%
%
%
%


\section{Cluster scale radio emission: new detections and candidates}

Table \ref{tab:radiodata} summarizes the results of the observations presented
in this paper. Cluster scale radio emission in the form of giant radio halo,
core--halo, peripheral emission was detected respectively in
A\,697, Z\,7160 and A\,781. Furthermore positive residuals, which
may suggest the presence of diffuse sources, were found
in A\,1682 and Z\,2661.
In this section we will present and briefly discuss these results.

\subsection{The giant radio halo in A\,697}

%
%
\begin{figure*}
\centering
\includegraphics[angle=0, width=18.5cm]{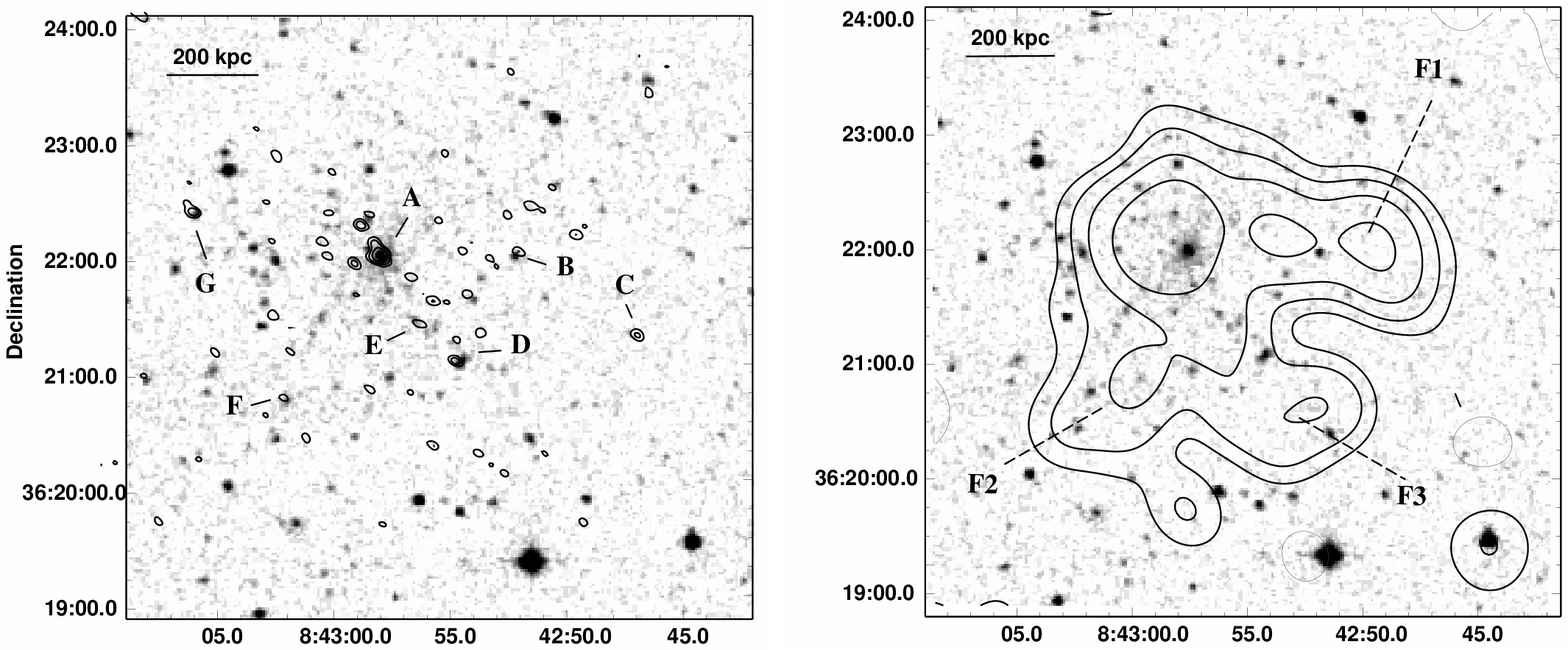}
\caption{{\it Left} -- 610 MHz GMRT radio contours of A\,697, overlaid on the 
DSS--2. The resolution of the image is 
$6^{\prime\prime} \times 5^{\prime\prime}$. The contour levels are 
$\pm$0.075,0.15,0.3,0.6,1.2,2.4,4.8,9.6 mJy b$^{-1}$. The rms level 
(1$\sigma$) is 25 $\mu$Jy b$^{-1}$. {\it Right} -- Same portion of the radio 
sky with a 610 MHz $40^{\prime\prime} \times 35^{\prime\prime}$ image overlaid.
The radio contours are $\pm$0.15,0.3,0.6,1.2,2.4,4.8,9.6 mJy b$^{-1}$. 
The rms level (1$\sigma$) is 50 $\mu$Jy b$^{-1}$. For this cluster 
1$^{\prime\prime}$ = 4.232 kpc.}
\label{fig:a697_ott}
\end{figure*}
%
%
No high sensitivity radio imaging is available in the literature
for A\,697 (z=0.2820, richness class 1, Bautz--Morgan type II--III).
Inspection of the cluster centre on the NRAO VLA Sky Survey 
(NVSS; Condon et al. \cite{condon98}) is not very insightful,
due to the presence of residual fringes in the image. Kempner \& Sarazin 
(\cite{kempner01}) reported the presence of candidate extended 
emission at the cluster centre on the basis of inspection of the 
327 MHz Westerbork Northern Sky Survey image (WENSS; Rengelink et al.
\cite{rengelink97}). 

An image of the central part of A\,697 at 610 MHz is shown in 
Fig. \ref{fig:a697_ott}, overlaid on the red optical frame from the 
Digitized Palomar Sky Survey (DSS--2). The full resolution image 
(left panel) shows a number of discrete
radio sources which are labelled from A to G. Source A is identified 
with the central cD galaxy; source B is associated with a backgroud 
galaxy, as derived from the redshift information in the 
Sloan Digital Sky Survey (SDSS); all the other sources have an optical 
counterpart at the redshift of the cluster according to the 
spectroscopic catalogue in Girardi et al. (2006) and to the SDSS. 
In addition to the individual radio galaxies, positive residuals 
of radio emission are clearly visible.
After subtraction of the 
flux density of the radio galaxies, which contribute for a total 
of 2.3 mJy, these residuals account for $\sim$9 mJy, suggesting 
the presence of low surface brightness emission at the cluster 
centre. 
\\
In order to highlight such emission, from the u--v data we subtracted all the
individual sources visible in the full resolution image, and produced an 
image with a resolution of $40^{\prime\prime} \times 35^{\prime\prime}$. 
This image is shown in the right panel of Fig.~\ref{fig:a697_ott},
and confirms the existence of extended emission in the form of a giant
radio halo, which extends on a largest angular 
scale LAS$\sim 3.5^{\prime}$, corresponding to a largest linear 
size LLS$\sim$ 890 kpc. Its total 
flux density, after subtraction of the individual radio galaxies (from A 
to G in left panel of Fig.~\ref{fig:a697_ott}) is S$_{\rm 610~MHz} = 13$ mJy, 
which implies a total radio power of P$_{\rm 610~MHz} = 3.5 \times 10^{24}$ 
W Hz$^{-1}$. The average surface brightness of the halo is of the order 
of $\sim 5\times 10^{-4}$ mJy arcsec$^{-2}$. 
This flux density value should be considered a lower limit. 
Indeed, on the basis of our analysis on the upper limits (Sect. 4) 
we estimate that a fraction of flux density of the order of $\sim$ 30 \% 
(from the outermost low brightness regions) may have been lost for this source.

The overall morphology of the giant radio halo is rather complex. 
The inner part of the source, 
i.e. a region of $\sim 1^{\prime}$ 
radius around the cluster centre, appears regular and symmetric. 
At larger radius a bright filament of emission (labelled as F1
in the right panel of Fig.~\ref{fig:a697_ott}) 
extends in the western direction, and two fainter features (F2 and F3) 
are located South of the cluster centre. The emission detected in the WENSS 
image (Kempner \& Sarazin \cite{kempner01})  appears elongated westward of 
the source peak. This elongation may correspond to the filament F1 observed 
at 610 MHz. The southern emission is  undetected in the WENSS image. 
However, we note that the very low brightness of the radio halo in the
outer parts does not allow a detailed comparison of
the fainter features.

An optical and X--ray analysis of A\,697 was carried out by Girardi 
et al. (2006), who found that the cluster is experiencing a very 
complex merging process. They argued that the dynamical state of A\,697 
might be explained either as an ongoing multiple accretion of small clumps by 
a very massive cluster, or as the result of a major merger occurring 
along the line of sight. This latter hypothesis would be consistent with 
the absence of a cool core in this cluster (Bauer et al. \cite{bauer05}). 

A more detailed discussion on the connection between the radio halo and the
cluster dynamics is in progress (Giacintucci et al. in prep.).

\subsection{The core--halo in Z\,7160}

Z\,7160 (z=0.2578) is a compact Zwicky Cluster (ZwCl\,1454.8+2223, 
NSCS\,J145715+222009), with little radio information available in
the literature. On the 1.4 GHz NVSS image the cluster centre is 
dominated by a compact source, coincident with the central galaxy. 
At the higher resolution of the ``Faint Images of the Radio Sky at
20 cm'' (FIRST) survey (White et al. \cite{white97}) 
the radio source is pointlike.
Chandra X--ray observations reveal that Z\,7160 is a cooling core cluster
(Bauer et al. \cite{bauer05}).

\begin{figure*}
\includegraphics[angle=0,width=18.5cm]{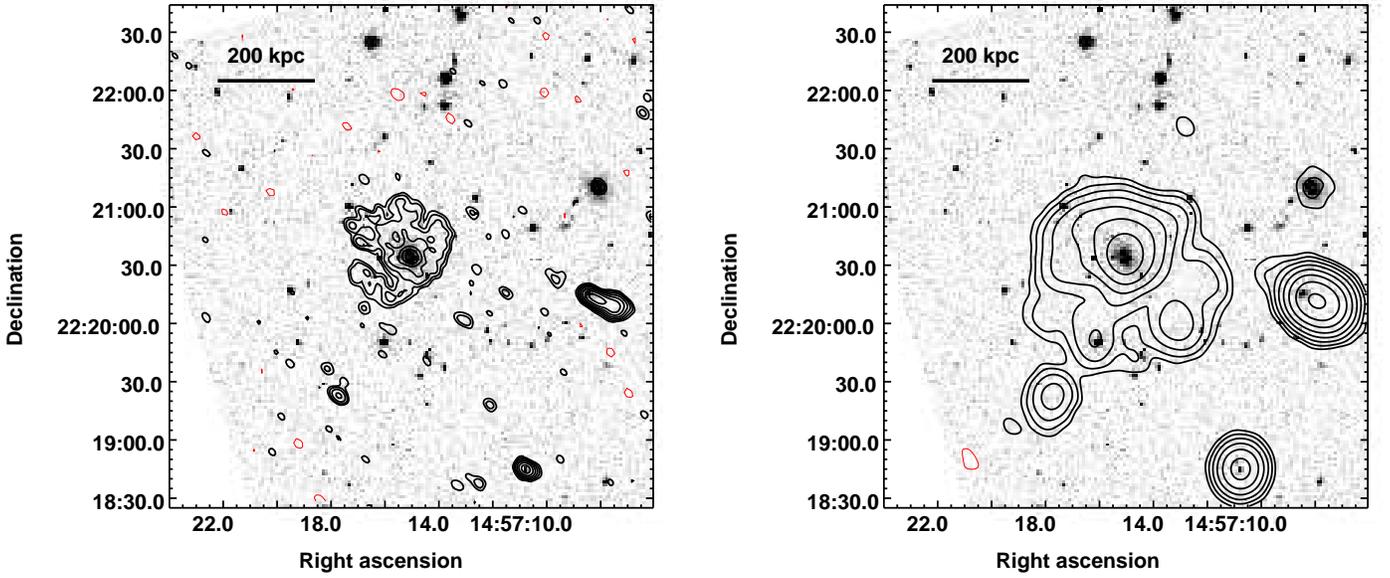}
\caption{{\it Left} -- 610 MHz GMRT radio contours of Z\,7160, overlaid on the 
optical red image from the SDSS. The resolution of the image is 
$5.9^{\prime\prime} \times 4.6^{\prime\prime}$. The contour levels are 
0.09 $\times$ ($\pm$ 1,2,4,8,16,32,64) mJy b$^{-1}$. The rms level (1$\sigma$) 
is 30 $\mu$Jy b$^{-1}$. {\it Right} -- Same portion of the optical sky with a 
610 MHz $17^{\prime\prime} \times 14^{\prime\prime}$ image overlaid. 
The radio contours are 0.12 $\times$ ($\pm$ 1,2,4,8,16,32,64,128) mJy b$^{-1}$.
The rms level (1$\sigma$) is 40 $\mu$Jy b$^{-1}$. For this cluster 
1$^{\prime\prime}$ = 3.965 kpc.}
\label{fig:z7160_ott}
\end{figure*}

The 610 MHz image of the cluster is reported in
Fig. \ref{fig:z7160_ott}, where a full resolution (left panel) and 
a tapered image (right panel) are shown. The central radio source 
consists of a compact component surrounded by an extended halo.
The total flux density of the source only slightly increases going from 
the  full resolution to the tapered image (from S=$38.0\pm1.9$ mJy to 
S=$43.6\pm2.2$ mJy), as well as the largest linear size, which goes 
from $\sim50^{\prime\prime}$ ($\sim$ 200 kpc) to $\sim90^{\prime\prime}$ 
($\sim$360 kpc). Both flux density values include the compact component
(associated with the dominant cluster galaxy) and the extended halo, whose 
origin is still debated (see for instance Mazzotta \& Giacintucci 
\cite{mg08}).
The radio contours in the full resolution image 
show that the source has a sharp edge, and the larger extent of the low
resolution image is the result of the convolution of positive residuals
on the western side of the source with a larger restoring beam.
In order to check if the extent of the central extended emission further
increases at decreasing resolution, we imaged the cluster at even lower 
resolutions, and obtained similar results in terms
of integrated flux density and largest linear size. We therefore trust
that the right panel of Fig. \ref{fig:z7160_ott} picks up all the
610 MHz radio emission.
\\
The morphology, size and radio power of this source, 
P$_{\rm 610~MHz}=8.25\times10^{24}$ W Hz$^{-1}$, coupled with its location in 
a cooling core, lead us to classify it as a core--halo source.
%
%

\subsection{Peripheral diffuse radio emission in A\,781}

\begin{figure*}
\centering
\includegraphics[angle=0,width=18.5cm,height=7.5cm]{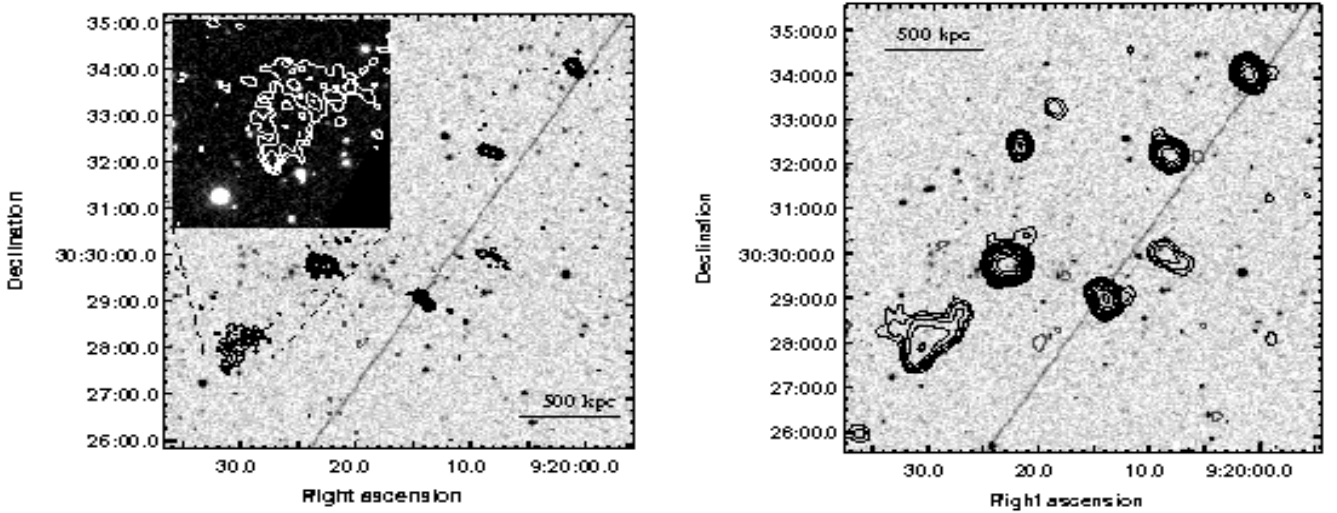}
\caption{{\it Left} -- 610 MHz GMRT radio contours of A\,781, overlaid on the 
DSS--2. The resolution of the image is $6^{\prime\prime} \times 
5^{\prime\prime}$.
The contour levels are $\pm$0.25,0.5,1,2,4,8,16 mJy b$^{-1}$. The rms level 
(1$\sigma$) is 50 $\mu$Jy b$^{-1}$. The insert zooms onto the 
peripheral radio source, superposed to the SDSS red optical frame. 
{\it Right} -- Same portion of the optical sky with a 610 MHz 
$18^{\prime\prime} \times 15^{\prime\prime}$ image overlaid. 
The radio contours are $\pm$0.3,0.6,1.2,2.4,4.8,9.6,19.2,38.4 mJy b$^{-1}$.
The rms level (1$\sigma$) is 50 $\mu$Jy b$^{-1}$. For this cluster 
1$^{\prime\prime}$ = 4.404 kpc.}
\label{fig:a781_ott}
\end{figure*}

A\,781 (z=0.2984, richness class 2, Bautz--Morgan type III) has been
studied very little at radio wavelengths. Its radio emission at 610 MHz is 
shown in Fig. \ref{fig:a781_ott}, overlaid on the red plate of the DSS--2. 
The cluster field is characterised by a number of extended
radio sources with optical counterpart, but the most outstanding feature  
is the extended low surface brightness emission peaked at 
RA$_{\rm J2000} = 09^h20^m32.2^s$, 
DEC$_{\rm J2000}=30^{\circ}27^{\prime}34.2^{\prime\prime}$. Hereinafter
we will refer to this source as to the ``peripheral radio emission''.
%
%
The insert in the left panel of Fig.~\ref{fig:a781_ott} shows that this 
source has no obvious optical counterpart, either in the proximity of the 
peak or underlying the radio emission. Its flux density is  
S$_{\rm 610~MHz} = 32\pm2$ mJy, which corresponds to a radio power
P$_{\rm 610~MHz} = 8.48\times10^{24}$ W Hz$^{-1}$ if we assume that it
is located at the cluster distance. At 1.4 GHz it is well visible on the  
NVSS, while it completely disappears on the FIRST survey (resolution of 
5 arcesc). This is a further indication of its very low surface brightness.
Using the NVSS flux density measurement we obtained a spectral index
$\alpha_{\rm 610~MHz}^{\rm 1.4~GHz} = 0.76$. This value is not as steep 
as it is usually  found in diffuse cluster sources and dying radio galaxies, 
where observations report values $\alpha$ \gtsim 1 (e.g. Giovannini \& Feretti 
\cite{gf04} and references therein for radio relics, and Parma et al.
\cite{parma07} for dying radio galaxies). 

\begin{figure*}
\centering
\includegraphics[angle=0,width=18cm]{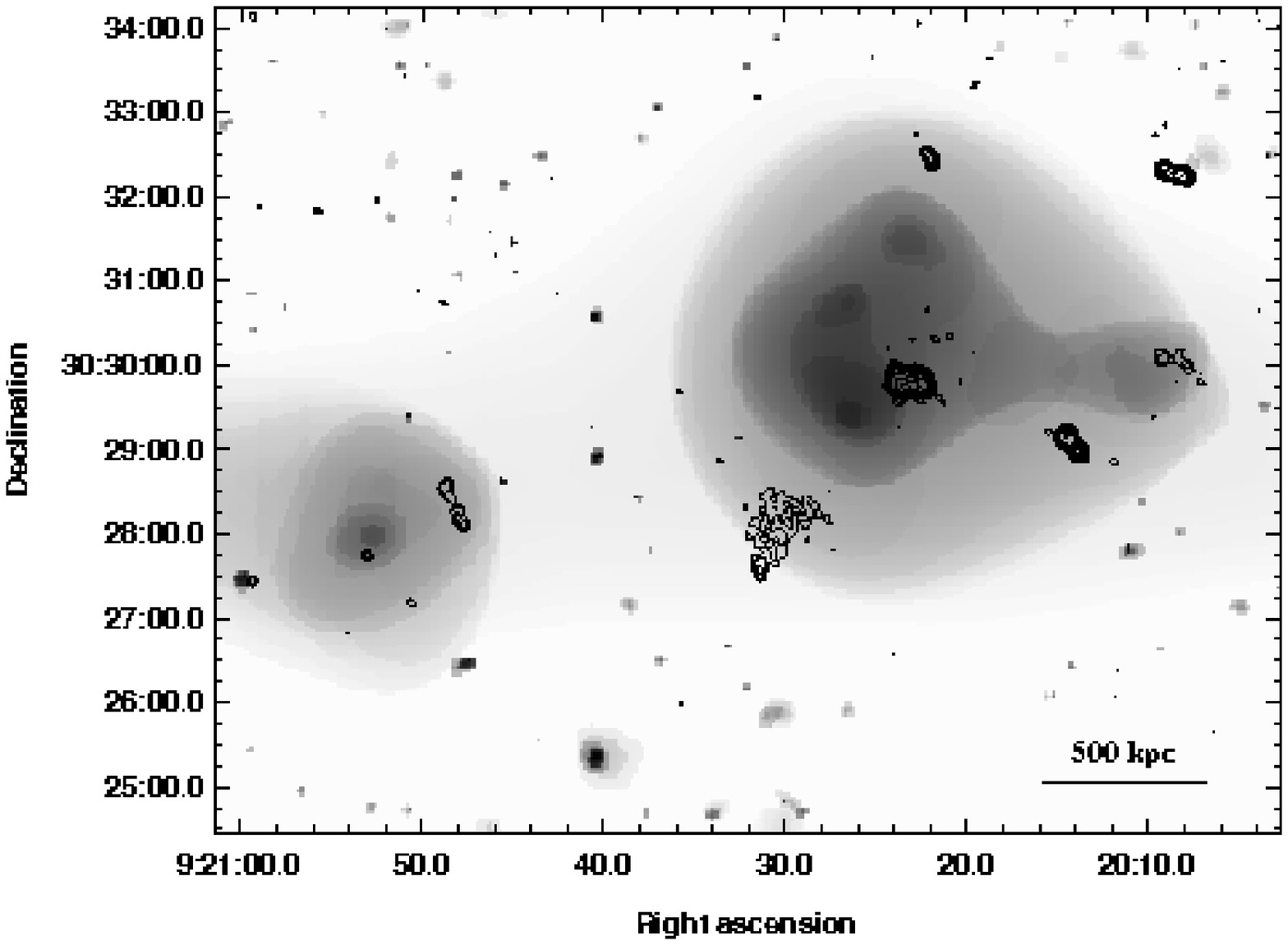}
\caption{610 MHz GMRT radio contours of A\,781, overlaid on the X--ray
wavelet--reconstructed Chandra image in the 0.5--5.0 keV band. The 
resolution of the radio image is $6^{\prime\prime} \times 
5^{\prime\prime}$. The rms level (1$\sigma$) is 50 $\mu$Jy b$^{-1}$.
The contour levels are $\pm$0.25,0.5,1,2,4,8,16 mJy b$^{-1}$.}
\label{fig:a781_x}
\end{figure*}

The cluster was observed with Chandra (Obs. Id. 534, ACIS--I, exposure 
$\sim$ 10 ks). 
The archival data were re--analysed by us in order to image the X--ray surface 
brigthness distribution, for comparison with the 610 MHz radio emission. 
Fig.~\ref{fig:a781_x} shows the wavelet--reconstructed image in the
0.5 -- 5.0 keV energy band, with 610 MHz radio contours overlaid.
%
%
The X--ray emission of the cluster is clearly complex.
Beyond the emission coming from the cluster centre, a secondary 
peak is located in the western direction, and embeds a tailed radio galaxy.
The peripheral radio emission is located South--West of the main X--ray clump 
detected by Chandra, and associated with the cluster centre. 
A third condensation of gas is visible South--East of the main X--ray
emission. An optical analysis from a lensing survey reveals that the S--E
X--ray condensation is coincident with another cluster at z=0.291, i.e. the
same distance of A\,781 (Geller et al. \cite{geller05}).
The information available in the optical band is not sufficient to 
perform a detailed analysis of the dynamical state of the cluster.

The origin of the peripheral radio emission is unclear. The lack of an optical
counterpart coupled with its location suggest that it might be 
a relic source, at the same time the value of the spectral index and its
overall radio morphology might suggest a tailed background radio source, whose
optical counterpart is too faint (or obscured) to be visible on the DSS--2.
This last possibility seems however implausible. No compact component is 
visible on the FIRST survey; moreover, the total radio power of this
source, if located at the average cluster distance, is already at the 
upper end 
for typical tailed radio galaxies. A 327 MHz GMRT follow up 
of this source is in progress, with the aim to throw a light on its origin.

\subsection{A\,1682 -- A giant radio halo and two relics?}

%
%
\begin{figure*}
\centering
\includegraphics[angle=0,width=18cm]{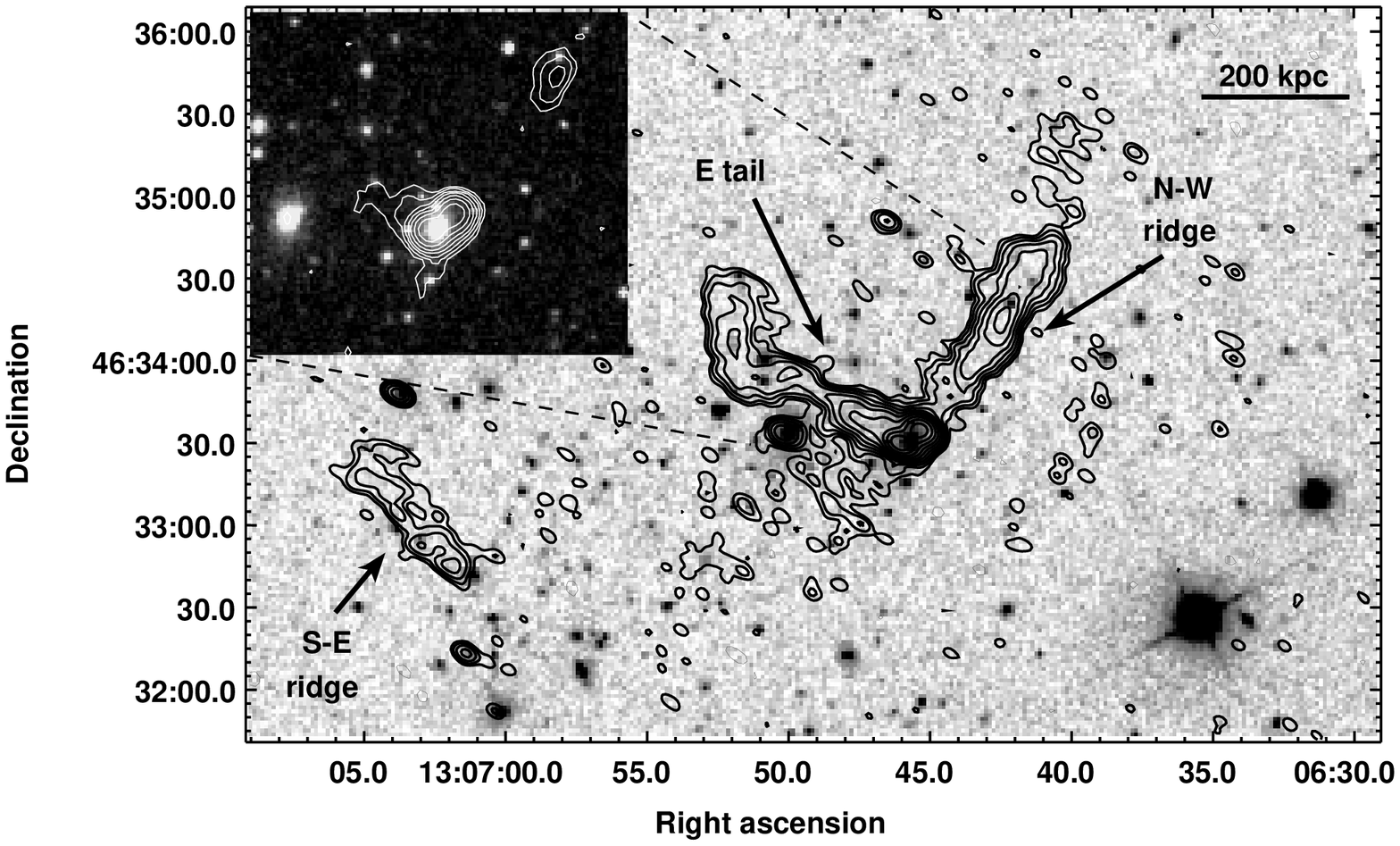}
\caption{610 MHz GMRT radio contours of A\,1682 on the optical red image
from the SDSS. The resolution of the image is 
$6.2^{\prime\prime} \times 4.1^{\prime\prime}$; 
the contour levels are 
0.12$\times~(\pm) 1,2,4,8,16,32,64,128,256,512$ mJy b$^{-1}$; 
the rms level (1$\sigma$) is 40 $\mu$Jy b$^{-1}$. The insert shows the 
1.4 GHz FIRST image of the cluster centre on the same optical frame. 
The resolution is $1.5^{\prime\prime}$. The radio contours are 
$\pm$0.45,0.9,1.8,3.6,7.2,14.4,28.8,57.6,115.2 mJy b$^{-1}$. For this cluster
1$^{\prime\prime}$ = 3.593 kpc.}
\label{fig:a1682_ott}
\end{figure*}

No high sensitivity and high resolution images are available in the
literature for A\,1682 (z=0.2260, richness class 1, Bautz--Morgan Type II).
The strong central radio emission visible at 1.4 GHz on the NVSS image 
turns out to be the convolution of discrete sources after inspection of 
the cluster centre on the FIRST image.

Our GMRT 610 MHz images are reported in Fig. \ref{fig:a1682_ott}.
The central radio emission in A\,1682 is extremely complex. 
The double--peaked radio emission is coincident with a cluster 
galaxy (z=0.2180) with a red magnitude (SDSS) $r=16.50$ 
(see insert).
Two tails depart from the inner region of the radio emission, however it 
is unclear if they are associated with the double radio galaxy. In the 
following we discriminate between ``the eastern tail'' and the ``north--western
ridge'' (see labels in the figure). While the eastern tail is might be
connected with the double--peakead source, the ridge (LAS $\sim 2^{\prime}$,
LLS $\sim$ 430 kpc) is more puzzling.
A possibility is that it is associated with another radio galaxy, however 
inspection of the radio--optical overlay shows that there is no obvious 
counterpart  responsible for the radio emission (see Fig. \ref{fig:a1682_ott}
and insert).
The lack of information at other radio frequencies does not allow 
any spectral analysis, which would be extremely useful to 
understand the nature of this ridge.
\\
Two more outstanding radio features are noticeable in the images displayed 
in Fig. \ref{fig:a1682_ott}: {\it (a)} the ridge of emission located 
South--East of the cluster centre (RA$_{\rm J2000} = 13^h05^m$,  
DEC$_{\rm J2000} = 33^{\circ}$), labelled as S--E ridge in the figure,
and {\it (b)} positive residuals spread all around the cluster centre. 

\begin{figure}
\centering
\includegraphics[angle=0,width=9.5cm]{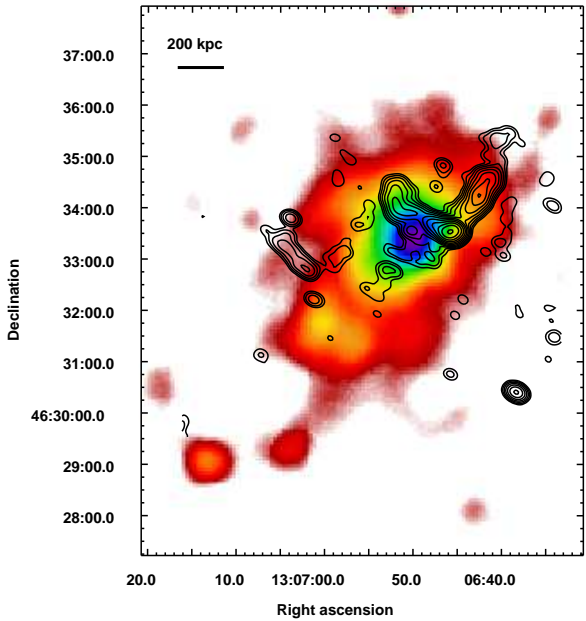}
\caption{610 MHz GMRT radio contours of A\,1682, overlaid on the 
X--ray smoothed Chandra image in the 0.5--5.0 keV band. 
The resolution of the radio image is $14.0^{\prime\prime} \times 
10.3^{\prime\prime}$. The rms level (1$\sigma$) is 50 $\mu$Jy b$^{-1}$.
The radio contours are 
$\pm0.2\times$(1,2,4,8,16,32,64,128,256,512,1024) mJy b$^{-1}$.}
\label{fig:a1682_x}
\end{figure}

\begin{itemize}

\item[{\it (a)}] The flux density of the S--E ridge is 
S$_{\rm 610~MHz} = 15.5\pm0.8$ mJy, its LAS is $\sim 1^{\prime}$ 
($\sim$ 215 kpc). This structure is visible also on the 
NVSS, where we measured a flux density S$_{\rm 1.4~GHz}$ = 6 mJy.
Due to the low resolution of the NVSS, this value includes also the 
contribution of the point source visible in Fig. \ref{fig:a1682_ott}
just north of the S--E ridge, whose flux density is  
S$_{\rm 610~MHz} = 3.2\pm0.2$ mJy. We can give a rough estimate of
the spectral index $\alpha_{\rm 610~MHz}^{\rm 1.4~GHz}$ of the S--E rigde
assuming that the point source has a $\alpha_{\rm 610~MHz}^{\rm 1.4~GHz}$=0.7,
which is reasonable for a compact source. This value for $\alpha$
provides S$_{\rm 1.4~GHz}$ = 1.8 mJy, and hence the estimated spectral index
for the ridge is $\alpha_{\rm 610~MHz}^{\rm 1.4~GHz} \sim$ 1.6.

\item[{\it (b)}] The presence of positive residuals spread over a region of 
$\sim 4^{\prime}$ around the cluster centre was very clear during the imaging 
process, however the presence of the central radio galaxy did not allow a 
proper imaging of this emission. Due to the complexity of the central 
radio emission, an accurate subtraction of the discrete sources
is not feasible. Nevertheless, we are confident that the 
positive residuals are significant. 
We estimated that the total flux density of the positive residuals, measured 
from the image shown in Fig. \ref{fig:a1682_x}, is up to 
S$_{\rm res}$(610 MHz)$\sim$ 44 mJy. 
This value was derived after subtrating the flux density of the discrete 
sources (the S--E and N--W ridges, the E--tail and the point sources) to the 
total flux density measured in a circular region with 
diameter of $\sim 4^{\prime}$. At the distance of A\,1682 this corresponds 
to a linear scale of $\sim$ 860 kpc. Note that the flux density remains fairly 
constant even integrating over a larger portion of the sky around the
cluster centre.

\end{itemize}

Figure \ref{fig:a1682_x} shows the 610 MHz radio emission of the cluster
center overlaid on the Chandra X--ray emission. The X--ray image is 
the result of re--analysis of archive data (Obs. Id. 3244, ACIS--I, exposure
$\sim$ 10 ks), carried out as for A\,781.
The gas distribution in A\,1682 is clearly disturbed. 
It shows two condensations, the north--western being considerably brighter 
and more extended than the south--eastern one. The positive residuals in the 
radio image, referred to in {\it (b)}, are all spread over the main X--ray 
condensation; the N--W ridge is located just outside the brightest
X--ray part, while the S--E ridge occupies a region of faint X--ray emission.

The galaxy distribution of the cluster, derived from a weak lensing analysis
(Dahle et al. \cite{dahle02}), confirms the perturbed dynamical state: 
two peaks and a third minor condensation are found at the location of the 
main gas condensation.

%
According to the paradigm that cluster mergers provide energy to produce
relativistic electrons,
the perturbed dynamical state in A\,1682 would be consistent with the 
hypothesis that it hosts a radio halo.
The residual flux density we measured\footnote{We note that the complex radio 
emission at the centre of A\,1682 makes this cluster very different from
those selected for the analysis on the upper limits. For this
reason no direct comparison can be made between this cluster and the flux 
density upper limits to the radio halo detection estimated in Section 4.}, 
S$_{\rm res}$(610 MHz)$\sim$ 44 mJy,
implies a radio power P$_{\rm 610~MHz} \sim 6.2\times10^{24}$ W Hz$^{-1}$.
This value would be in agreement with the 
logP$_{\rm 1.4~GHz}$ -- logL$_{\rm X}$ correlation shown in a number of papers
(e.g. Liang et al. \cite{liang00}; CBS06) 
and scaled assuming $\alpha_{\rm 610~MHz}^{\rm 1.4~GHz}$=1.3 (i.e. Brunetti
et al. \cite{brunetti07}).
The overall appearance of the cluster, and the lack of an optical 
counterpart for the S--E and N--W ridges, suggest that these regions of 
radio emission might be connected with merging processes, and
one possibility is that they are relic sources.
Low frequency radio follow up for this cluster is in progress and a more
detailed analysis will be presented in a forthcoming paper.

\subsection{Candidate extended emission at the centre of Z\,2661}

Z\,2661 is a compact Zwicky Cluster (Zw\,0947.2+1723), the most X--ray 
luminous and the second most distant in the sample, at z=0.3825
(see Tab.~1). 
Very little radio and X--ray information exists in the literature. 
Chandra X--ray observations are available from the archive, and 1.4 GHz 
images are available from the NVSS and FIRST
surveys. An elongated diffuse source is visible on the NVSS in the central 
part of the cluster, with a largest linear size $\sim3^{\prime}$. Inspection 
of the FIRST image shows that only a faint point--like source is located 
within the extended emission on the NVSS.
%
%
\begin{figure*}
\centering
\hspace{-1cm}\includegraphics[angle=0,width=17cm]{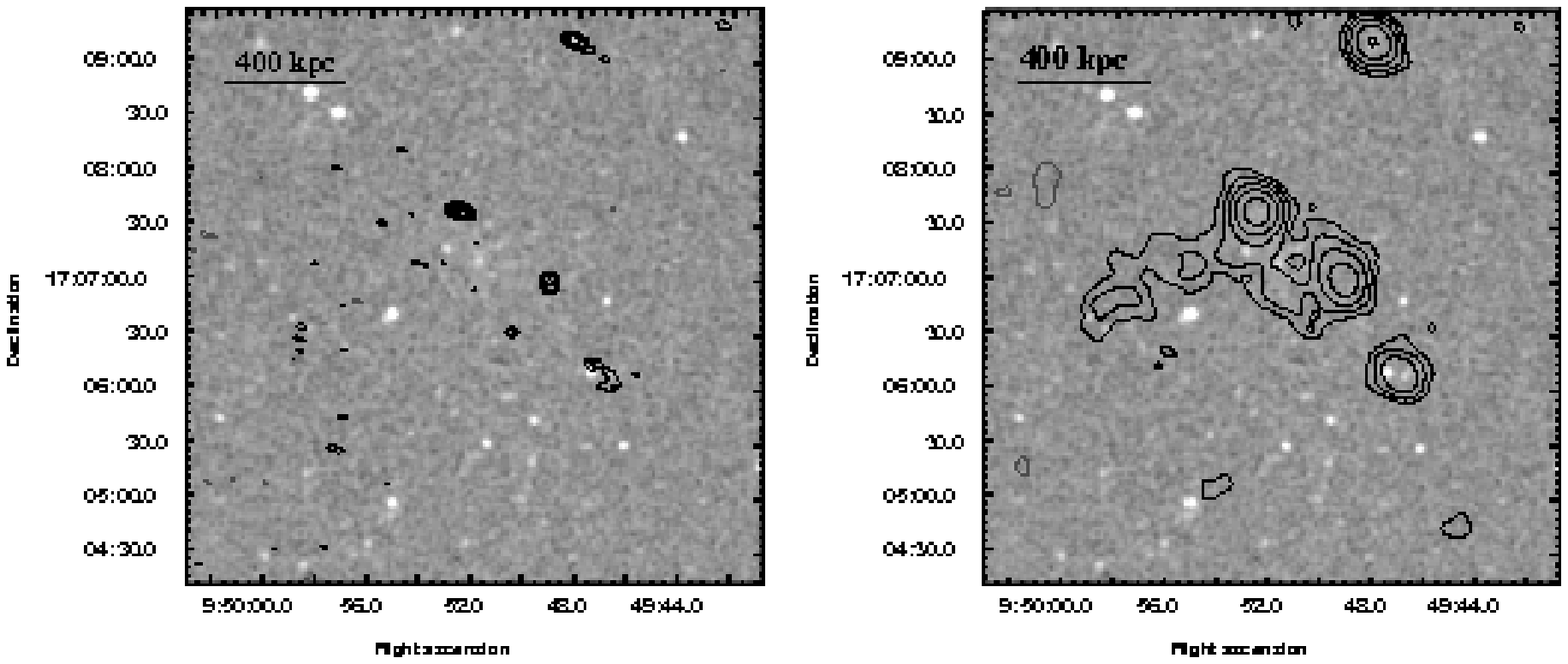}
\caption{{\it Left} -- 610 MHz GMRT radio contours of Z\,2661, overlaid on 
the DSS--2. The resolution of the image is 
$6.5^{\prime\prime} \times 4.5^{\prime\prime}$, p.a. 
80$^{\circ}$. The contour levels are $\pm$0.25,0.5,1,2,4,8,16 mJy b$^{-1}$. 
The rms level 
(1$\sigma$) is 65 $\mu$Jy b$^{-1}$. {\it Right} -- Same portion of the radio 
sky with a 610 MHz $17.7^{\prime\prime} \times 17.4^{\prime\prime}$ image 
overlaid. The radio contours are $\pm$0.25,0.5,1,2,4,8 mJy b$^{-1}$. 
The rms level (1$\sigma$) is 65 $\mu$Jy b$^{-1}$. For this cluster 
1$^{\prime\prime}$ = 5.196 kpc.}
\label{fig:z2661_ott}
\end{figure*}

\begin{figure}
\centering
\includegraphics[angle=0,width=9cm]{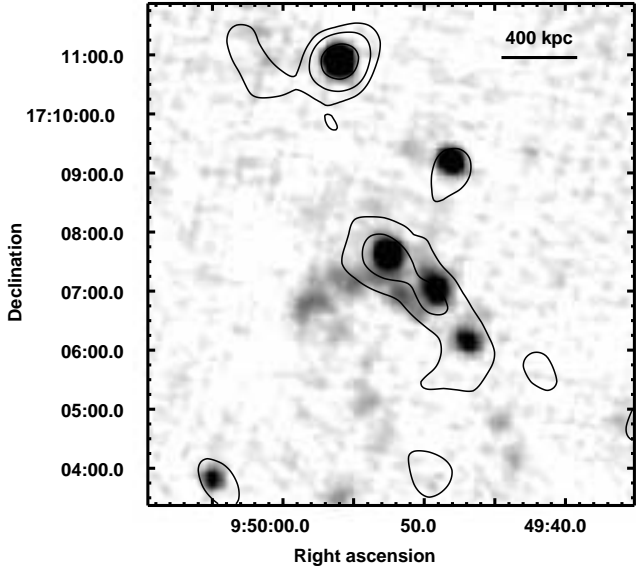}
\caption{Grey scale GMRT 610 MHz emission of Z\,2661 (same image as in the 
right panel of Fig.~\ref{fig:z2661_ott}), with 1.4 GHz NVSS contours 
overlaid. The contour levels are $\pm$1,2,4 mJy b$^{-1}$.}
\label{fig:gmrt_nvss}
\end{figure}

\begin{figure}
\centering
\includegraphics[angle=0,width=9cm]{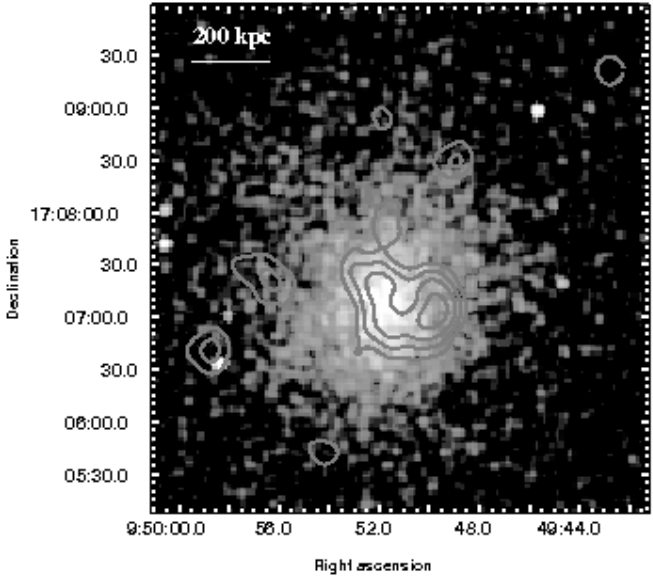}
\caption{610 MHz GMRT contours of the central residual emission
in Z\,2661, overlaid on the smoothed Chandra image
in the 0.5--5.0 keV energy band. The restoring beam is
$20^{\prime\prime}\times20^{\prime\prime}$. Contours are 
$\pm$ 0.15,0.3,0.5,0.8 mJy b$^{-1}$. The average rms close to the
central residuals is $\sim~40~\mu$Jy b$^{-1}$.}
\label{fig:z2661_ext}
\end{figure}

The radio emission of Z\,2661 at 610 MHz is shown in Fig.~\ref{fig:z2661_ott},
where a full resolution (left panel) and tapered image (right panel)
are shown overlaid on the DSS--2.  The most relevant 
feature in the $17^{\prime\prime}\times17^{\prime\prime}$ resolution image
(right panel) is the extended emission visible around the three discrete
radio sources in the central cluster region, partially coincident with 
the diffuse emission visible on the NVSS, as shown in 
Fig.~\ref{fig:gmrt_nvss}.
In order to highlight this emission, from the u--v data we subtracted all the 
discrete sources visible in the full resolution image of the whole cluster
field,
and produced an image of the residuals using only the shortest baselines and 
natural weighting. The resulting image is shown in Fig.~\ref{fig:z2661_ext}, 
superposed on the Chandra X--ray image, which was obtained from a 
re--analysis of data available from the 
public archive (Obs. Id. 3274, ACIS--I, exposure $\sim$ 15 ks).
\\ 
The X--ray morphology of Z\,2661 looks only marginally 
perturbed, however a change in position angle of the isodensity colours
with changing distance from the cluster centre suggests that projection 
effects might play a role on the overall appearance of the brightness
distribution. 
The radio--X overlay 
clearly shows that the extended residual emission is located on the peak of 
the X--ray surface brightness of the intracluster gas. The spatial 
coincidence between the thermal and non--thermal emission is a further hint 
that the residuals might belong to a diffuse cluster radio source.

The total flux density of this emission ranges from $\sim$ 4 to 
$\sim$ 5.9 mJy, depending on the size of the region considered for the
integration. Based on our analysis on the upper limits described in
Sec.~\ref{sec:uplim}, such residual values are 
expected in the case of radio halos with total injected flux density
$S_{\rm inj} \approx 10 \div 15$ mJy. We thus consider this case 
a ``suspect'' diffuse emission.
Using the measured flux density, the radio power of the extended emission 
at the centre of Z\,2661 would be in the range 
P$_{\rm 610~MHz} \sim 2 - 3 \times 10^{24}$ W Hz$^{-1}$; 
the angular size of the imaged emission is $\sim 1^{\prime}$ ($\sim$ 310 kpc).
We stress however that these values are very uncertain and refer only
to the information provided by the images.
Radio observations at lower frequency are necessary to confirm and 
further study this emission.
\\
\\

\subsection{Notes on the clusters with VLA archive data}

\begin{figure}
\centering
\includegraphics[angle=0,width=9cm]{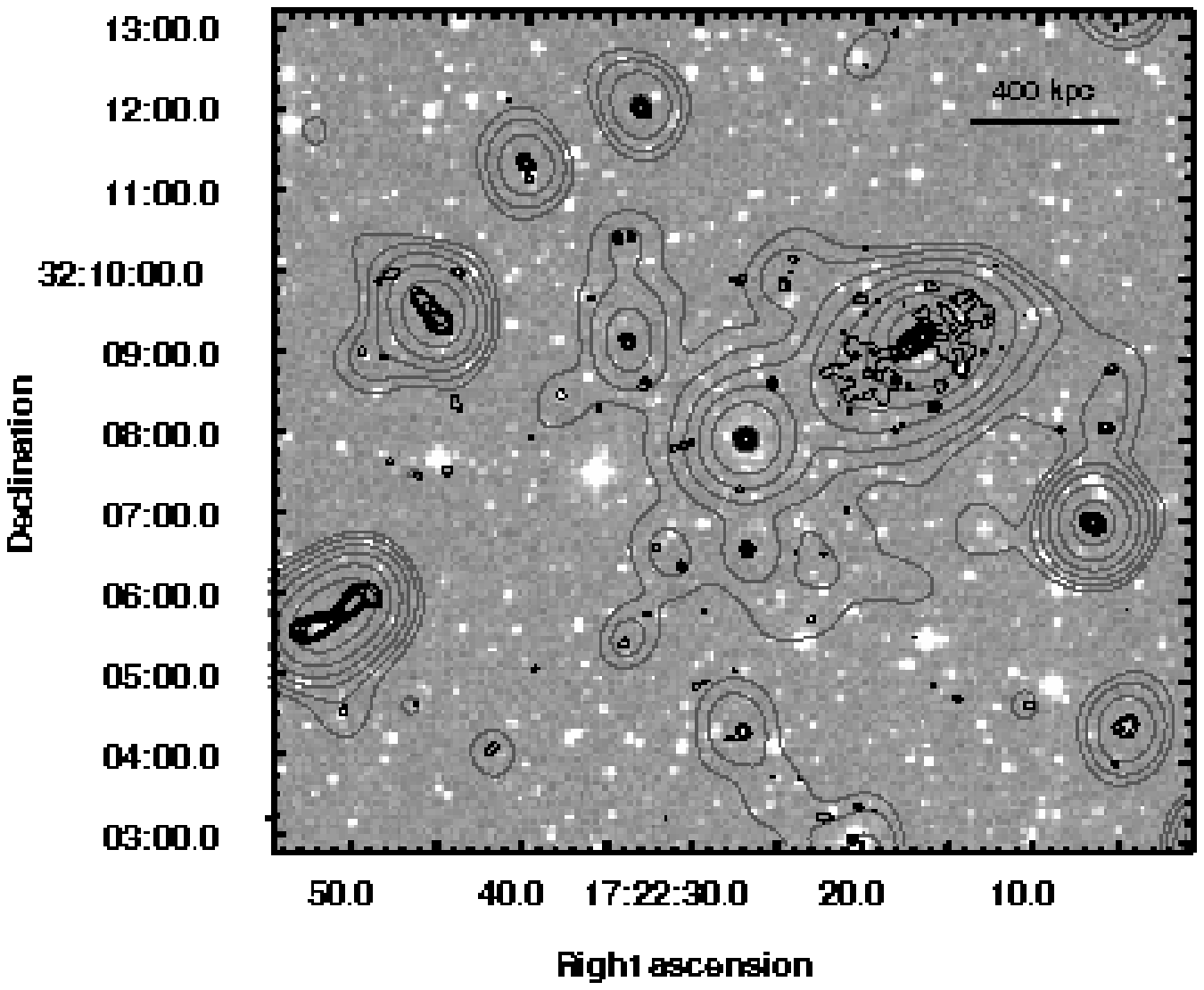}
\caption{1.4 GHz emission in A\,2261 overlaid on the DSS--2 optical image.
Red contours: D--array, resolution of 
$44.6^{\prime\prime}\times41.3^{\prime\prime}$, contour levels $\pm$ 0.15,
0.3, 0.6 ... mJy b$^{-1}$, 1$\sigma$ rms 50$\mu$J b$^{-1}$.
Black contours: B+D array, resolution 
$6.3^{\prime\prime}\times5.7^{\prime\prime}$, contour levels $\pm$ 0.09,
0.18, 0.36 ... mJy b$^{-1}$, 1$\sigma$ rms 30$\mu$J b$^{-1}$. For this cluster
1$^{\prime\prime}$=3.568 kpc.}
\label{fig:a2261_ott}
\end{figure}

A number of clusters belonging to the sample in Table 1 were not observed 
as part of the GMRT radio halo survey either because they belong to other
GMRT programs (i.e. the GMRT cluster key project), or because of the existence 
of VLA archival data.
\\
We retrieved and re--analysed 1.4 GHz VLA D-- configuration archival data of 
A\,1763 (Obs. Id. AC696), A\,2111 (Obs. Id. AG639), A\,2261 (Obs. Id. AC696). 
No hint of large scale emission was found in A\,1763 and A\,2111.
The case of A\,2261 is more uncertain, due to the presence of strong and 
extended sources at the cluster centre. In the attempt to further investigate 
the nature of the central emission in this cluster, we re--analysed also 1.4 
GHz VLA B--configuration archival data (Obs. Id. AC696) and combined the two 
arrays. 
The resulting images are shown in Fig. \ref{fig:a2261_ott}, where
the radio contours of the D (red) and  B+D (black) arrays are reported, 
overlaid on the DSS--2. Even though a detailed analysis is not possible,
due to some problems with the flux density scale of the two arrays,
our images show that some extended emission might be present around the 
dominant galaxy over an angular scale of 
4$^{\prime} - 5^{\prime}$ ($\sim 850 - 1000$ kpc). An accurate
subtraction of the contribution from the discrete sources (unfeasible with
the available archival data) is needed to clarify the situation.
%
%

\section{Discussion of the GMRT radio halo survey observational results}

We summarize the most important results reported in this 
work, and give a brief discussion of the results of the whole sample,
including the literature information and those presented in Paper I.
We remind here that no information is available in the literature for the 
five clusters marked with $\surd$ in Table 1.
An overview of the diffuse cluster sources found in the GMRT radio halo
survey is reported in Table \ref{tab:summary}.

%
\begin{table*}[t] 
\caption[]{Results of the GMRT Radio Halo Survey. Halos, relics and candidates}
\begin{center}
\begin{tabular}{lccclrl}
\hline\noalign{\smallskip}
Cluster & Source Type & S$_{\rm 610~MHz}$ & logP$_{\rm 610~MHz}$ &  LAS & LLS & Axial Ratio    \\ 
        &             & mJy               &  W Hz$^{-1}$         & arcmin & kpc &     \\
\noalign{\smallskip}
\hline\noalign{\smallskip}
A\,209$^{\star}$ & Giant Halo & 24.0 $\pm$ 3.6 & 24.46 & $\sim$ 4 & $\sim$  810 & $\sim$ 2   \\
RXCJ\,2003.5--2323$^{\star}$  & Giant Halo  &   96.9 $\pm$ 5.0  &  25.49  & $\sim$ 5   & $\sim$ 1400 & $\sim$ 1.3 \\
A\,697           & Giant Halo & 13.0 $\pm$ 2.0 & 24.54 & $\sim$ 3.5 & $\sim$ 890& $\sim 1$   \\
RXCJ\,1314.4--2515$^{\star}$  & Western Relic &   64.8 $\pm$ 3.2 &  25.03 & $\sim$ 4   & $\sim$  910 & $\sim$ 3 \\
                  & Eastern Relic & 28.0 $\pm$ 1.4 & 24.67 & $\sim$ 4   & $\sim$  910 & $\sim$ 4.3 \\
                  & Halo          & 10.3 $\pm$ 0.3  & 24.22  & $\sim$ 2 & $\sim$  460 & $\sim$ 1.5 \\
A\,521$^{\star}$ & Relic      & 41.9 $\pm$ 2.1 & 24.91 & $\sim$ 4 & $\sim$  930 & $\sim$ 4.5 \\ 
A\,3444$^{\star}$   & Central Galaxy        &   16.5 $\pm$ 0.8  &  24.51  & $\sim$0.7  & $\sim$  165 & $\sim$ 1.4  \\
          & surrounding Halo   & 10.0 $\pm$ 0.8 & 24.29 & $\sim$1.5  & $\sim$ 350 & $\sim$ 1.4  \\
Z\,7160           & Core--halo & 43.6 $\pm$ 2.2 & 24.92 & $\sim$ 1.5 & $\sim$ 360 & $\sim$ 1 \\
A\,781    & Candidate relic?   & 32.0 $\pm$ 2.0 & 24.92 & $\sim$ 2   & $\sim$ 520 & $\sim$ 1.7 \\
A\,1682$^{\diamondsuit}$ & Candidate halo? & $\sim$ 44      & 24.79 & $\sim$ 4   &    --     & -- \\
Z\,2661$^{\clubsuit}$ & Candidate halo?    & $\sim$ 5.9     & 24.47 &     --     &    --    & -- \\
\hline{\smallskip}
\end{tabular}
\end{center}
$^{\star}$ Information from Paper I.\\
$^{\diamondsuit}$ See Section 5.4.\\
$^{\clubsuit}$ See Section 5.5.
\label{tab:summary}
\end{table*}

\subsection{Cluster centres} 
\subsubsection{Radio halos and candidates}

Among the clusters presented in this paper, a new giant (LLS $\sim$ 890 kpc) 
radio halo was discovered in A\,697. 
This brings to 10 the total number of clusters with a radio halo 
in our selected sample. In particular: 
A\,2744, A\,1300, A\,2163, A\,773, A\,1758a and A\,2219 have
literature information (see Table 1); A\,209, A\,697, RXCJ\,1314.4--2515 and 
RXCJ\,2003--2323 were found with the GMRT radio halo survey. 
They are all giant halos, except RXCJ1314.4--2515 
(see Table \ref{tab:summary}). 

A candidate giant radio halo was found in A\,1682 (Section 5.4). We found
residual emission of $\sim$ 44 mJy, spread over a linear scale of the order
of $\sim$ 860 kpc. The very unrelaxed morphology of the cluster would
argue in favour of the presence of a giant radio halo, which we were
unable to properly image with the 610 MHz data due to the highly complex 
radio emission at the cluster centre. Follow up observations are in progress 
to confirm this result. 

``Suspect'' extended emission was found at the centre of Z\,2661 (Section 5.5),
the second most distant and second most X--ray luminous cluster in the
sample. It is possible that we are imaging only the peak brightness of a 
brighter and larger structure. 
Follow up observations at lower frequencies are necessary to confirm 
this result.

The re--analysis of archival VLA data shows that the central region of A\,2261 
(Section 5.6) is complex and puzzling, due to the presence of many individual 
sources and to the very extended emission associated with a cluster FRI galaxy 
(Fanaroff \& Riley \cite{fr74}). Further investigation is 
necessary to disentangle the contribution of the individual sources from 
possible extended emission on the cluster scale.

Finally, three  clusters in the sample host central extendend radio 
emission, though of different origin. The nature of the central radio emission 
in A\,2390 (Bacchi et al. \cite{bacchi03}) is still debated and it is
unclear if it is a mini--halo source (see Gitti et al. \cite{gitti04} for 
details on this class of extended radio sources) or a radio halo of small 
size; a core--halo source was found in the cool core cluster Z\,7160 
(Section 5.2), and candidate extended emission, possibly associated with the 
dominant cluster galaxy, was found in A\,3444 (Paper I).

\subsubsection{Non detections and upper limits}

Our work confirms that radio halos are a rare phenomenon. One of the main
results presented here and in Paper I is that most of the clusters in our 
sample do not show central extended emission. 

Among the 34 clusters observed with the GMRT radio halo survey, 25 lack
a central extended source at the sensitivity level of the observations.
For 20 of them we derived firm upper limits to the radio power of a
giant radio halo by means of a radio analysis carried out on typical GMRT u--v 
data sets and using the rms noise in the final images (Section 4). 
For the remaining five clusters we did not derive the upper limits, since
the rms noise in the final images is considerably higher than the average
quality of our survey (see Table 2). Such clusters will not be considered
in Section 6.4.
\\
Furthermore, no emission was found with the VLA in RXCJ\,0437.1+0043 
(Feretti et al. 2005), in A\,1763 and A\,2111, whose archival data were 
re--analysed here (Section 5.6).

\subsection{Cluster outskirts. Radio relics and candidates}

The 610 MHz GMRT radio halo survey turned out to be successful also for 
the study of the peripheral cluster regions, where relic sources are
usually found. Our knowledge on the origin of radio
relics is still limited. At this stage, a clear correlation exists between the
presence of relics and dynamical activity in the hosting clusters, and
all models proposed so far to explain the origin of relics invoke
the presence of a merger shock (e.g. En\ss{}lin et al. \cite{ensslin98};
Roettiger et al. \cite{roettiger99};
En\ss{}lin \& Gopal--Krishna \cite{egk01}; Markevitch et al. \cite{mm05}).
However, only few relics have been studied in detail so far, 
and new detections of relic sources are crucial to improve our
understanding on the origin of these sources.

We found a relic in A\,521 (Paper I; Giacintucci et al.  \cite{g06} 
and \cite{g08}) and a double relic in RXCJ\,1314.4--2515 
(Paper I; Feretti et al. \cite{feretti05}). RXCJ\,1314.4--2515
is a unique case of a cluster with a double relic and a radio halo.
We are studying both clusters over a wide range of radio frequencies.

Two more clusters deserve further 
investigation. Peripheral extended emission, whose origin is
unclear, was detected in A\,781 (Section 5.3); two intriguing ``rigdes'' 
of radio emission were found in  A\,1682, elongated in shape and with no 
optical counterpart (Section 5.4). The X--ray information on both
clusters suggests a complex dynamics.

\subsection{Information on the cluster dynamics from X--ray data}

It is fairly well established that all clusters hosting halos and relics
show signs of merging activity (e.g. Buote \cite{buote01}; 
Schuecker et al. \cite{schuecker01}), while a complete analysis of the 
dynamics of 
clusters without halos and relics is still missing. A huge amount of data 
would be necessary to perform such study in the optical band, and our main 
source of knowledge is the information of the gas distribution available from
the X--ray data archives.
\\
In order to collect some information on the dynamical state for each cluster 
in our sample, based either on the mass and/or 
intracluster gas distribution, we searched into the literature
(Dahle et al. \cite{dahle02}; Zhang et al. \cite{zhang06}). Furthermore
we consulted the X--ray public archives (ROSAT, Chandra, ASCA and XMM) 
and carried out a zero--order data reduction for imaging purposes (i.e. no 
data analysis) for all clusters (regardless of their radio
properties) without literature information and with public observations 
in the Chandra and/or XMM archive.
\\
\\
Our investigation reinforces our earlier findings and the literature 
information:  all clusters hosting halos and/or 
relics and those with candidate extended emission show signatures 
of dynamical activity, i.e. asymmetries and substructure in the X--ray 
emission, elongation and/or multiple peaks in the distribution of the mass
(see Paper I, references in Table 1 and Sections 5.1, 5.3, 5.4 and 5.5).

24 clusters in our complete sample do not host diffuse extended
emission of any kind. This number includes 19 upper limits in Table 3 
(we dropped RXCJ\,2228.6+2037 due to the redshift), A\,1763 and A\,2111
(Section 5.6), RXCJ\,0437.1+0043 (Feretti et al. \cite{feretti05}), and the
two clusters A\,3444 (Paper I) and Z\,7160 (Section 5.2), whose central
emission seems connected to the central galaxy.
No information was found for three clusters (Z\,5768, RXCJ\,1512.2--2254 and 
Z\,5699). From a visual inspection of the X--ray images for the remaining 21
we found that:

\begin{itemize}

\item[{\it(a)}] seven clusters show a fairly regular gas distribution, 
and they do not seem to be undergoing dynamical activity;

\item[{\it(b)}]  14 clusters show irregular distribution 
of the intracluster gas, in the form of elongation, multiple clumps, 
change of boxiness at different cluster radii. These morphologies are 
suggestive of  some form of dynamical activity;

\item[{\it (c)}] if we divide our clusters into two bins of X--ray luminosity
and define a ``high'' luminosity and a ``low'' luminosity interval, 
with the separation at 
L$_{\rm X} \approx 8\times10^{44}$ erg s$^{-1}$ (see also 
Cassano et al. \cite{cassano08}), 
regular and disturbed  clusters are equally found in both intervals.

\end{itemize}

\begin{figure}
\centering
\includegraphics[angle=0,width=9cm]{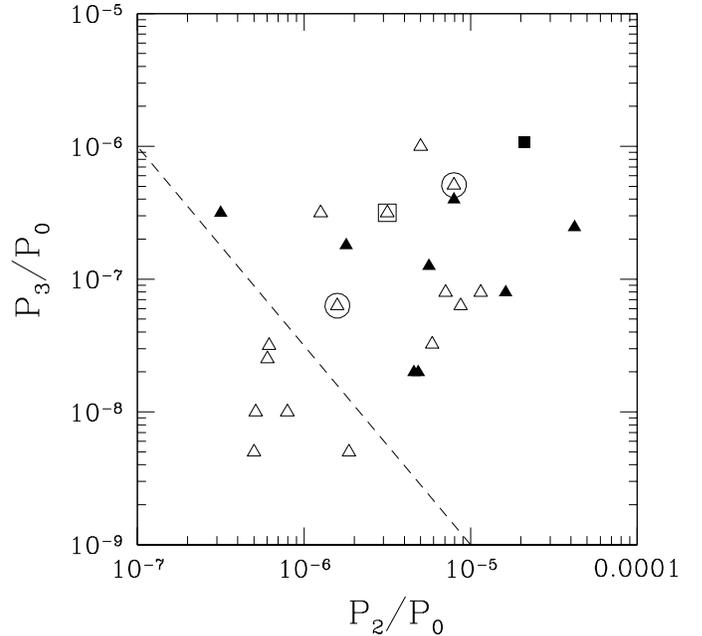}
\caption{Location of clusters with and without extended cluster scale 
emission on a P$_2$/P$_0$ -- P$_3$/P$_0$ plot. Filled triangles:
clusters with radio halo; empty triangles: clusters without radio halo; 
filled square: cluster with radio relic; empty triangles inside a
circle: clusters with candidate halo; empty triangle inside a square:
clusters with candidate relic. The dotted line divides the relaxed
and perturbed clusters.}
\label{fig:pratio}
\end{figure}

Our qualitative findings are corroborated by a comparison of 
the results for our sample and the work of Hart (\cite{hart08}),
who studied the X--ray morphology for a sample of 143 
galaxy clusters with available Chandra public observations
by means of a  multiple--moment power ratio analysis 
(Buote \& Tsai \cite{bt95}).
\\
A number of clusters in our sample are included in the X--ray analysis
by Hart 
and for them we could cross--correlate the radio properties (presence of
cluster diffuse emission or lack thereof) with a quantitative indicator
of the distribution of the X--ray intracluster gas. 
The results are reported in Fig. \ref{fig:pratio}, which shows the location 
of the clusters in the P$_2$/P$_0$ -- P$_3$/P$_0$ plane. Each
P$_n$ represents the square of the $n$th multipole of the two--dimensional
pseudopotential generated by the X--ray surface brightness evaluated
over a circular aperture. The ratio
P$_2$/P$_0$ is a measurement of the ellipticity, and P$_3$/P$_0$ is
related to the presence of multiple peaks in the X--ray distribution
(see Buote \cite{buote01} and Hart \cite{hart08} for details on these 
parameters).
Low values both of P$_2$/P$_0$ and of P$_3$/P$_0$ indicate little deviation 
from spherical symmetry and may be considered indicators of relaxation; 
the dotted line is a tentative separation between small and large deviations 
from symmetry.
\\
Some interesting conclusions can be drawn from the figure. The 
clusters below the dotted line may be considered relaxed clusters, and
none of them hosts diffuse cluster emission of any kind; those above the
line, which show large deviations from symmetry and may be considered
perturbed clusters, include radio halos, relics, candidates for both 
classes, as well as clusters without diffuse radio sources.

%
%

\subsection{Fraction of clusters hosting a radio halo}
 
A preliminary estimate of $f_{\rm RH}$, the fraction of clusters in 
our sample which host a radio halo, can finally be given. 
Unfortunately the radio information is
not complete for all clusters in the sample, so in the normalization 
we cannot include the 5 clusters marked with $^{\surd}$ in Table 1.  
We further exclude the 5 clusters with high rms (see Section 4), those with 
suspect central diffuse emission (A\,521, A\,2261, Z\,2661 and A\,1682, see 
Section 4 and 6.1.1), and the cluster RXCJ\,2228.6+2037, whose redshift is
just above the limit of our selection criteria (see Table 1). 
Therefore the fraction is $f_{\rm RH}$ 10/35 = 29$\pm9$\%.
\\
Most of the clusters with radio halo, i.e. 9/10, are found in the high 
X--ray luminosity bin, where the fraction of clusters with radio halo 
is as high as $f_{\rm RH}$ 9/24 = 38$\pm13$\%. We point out that
the uncertainty in $f_{\rm RH}$ is derived on the basis of the detections 
only.
\\
Giovannini et al. (\cite{giovannini99}) cross--correlated the X--ray Brightest 
Abell clusters (XBACs sample) with the NVSS for the local Universe 
(z$\le$ 0.2) and found that the occurrence of cluster halos and relics is 
higher in clusters with high X--ray luminosity and high temperature.
Our result confirms those findings,
i.e. also at higher redshift clusters with high X--ray luminosity have a 
higher probability to host a radio halo. A detailed analysis of the
statistics on the occurrence of radio halos over the redshift range
0.05 -- 0.4 was carried out in Cassano et al. (\cite{cassano08}).

\section{Summary and Conclusions}

The GMRT radio halo survey, carried out at 610 MHz, was designed in
order to investigate the link between the presence of diffuse cluster scale 
emission and the dynamics of the hosting clusters in the redshift
interval 0.2 -- 0.4, where in the framework of the re--acceleration model 
the bulk of radio halos is expected. 
To this aim we selected a complete sample of 
massive clusters with L$_{\rm X}$(0.1--2.4 keV) $>$ 5 $\times$ 10$^{44}$ 
erg s$^{-1}$, and $0.2<{\rm z} \le0.4$ from the ROSAT-ESO Flux Limited X--ray 
galaxy cluster catalogue and from the extended ROSAT Brightest Cluster
Sample. Declination limits were also imposed (Section 2).

34 clusters in the sample lacked high sensitivity information in the
literature and in the radio archives, and were observed with the GMRT at 
610 MHz. Imaging was carried out over a wide range of resolutions, 
with sensitivity in the average range 35 -- 100 $\mu$Jy b$^{-1}$. 
We also retrieved and analyzed unpublished archive VLA observations 
for three clusters in the sample. 11 GMRT clusters were 
presented in Paper I, the remaining are presented in this paper. The main
results of the GMRT radio halo survey can be summarized as follows.

\begin{itemize}

\item[(1)]
Diffuse emission on the cluster scale was found in a number of clusters.
In particular, we detected: 4 radio halos, bringing to 10 their number 
in the whole sample; three relic sources; one core--halo
cluster. Moreover, we found a candidate giant radio halo, suspect emission
at the centre of one cluster, a candidate relic and a candidate core--halo
source. For all clusters with candidate extended emission follow--up
observations are in progress. 

\item[(2)]
Most of the clusters observed with the GMRT do not host diffuse 
central emission, and this result confirms that radio halos are not common. 
Thanks to the high quality of our data we managed to 
place firm radio power upper limits in 20 clusters.

\item[(3)]
For those clusters with available X--ray information, we carried
out a X--ray/radio analysis.
Our results confirm that all clusters hosting halos, relics and candidate
diffuse emission show signature of dynamical activity. Furthermore, we 
find that clusters without diffuse emission may be both relaxed and 
perturbed. Finally, none of the relaxed clusters hosts diffuse emission.
A more quantitative analysis would be required in order to assess the 
amount of turbulent energy which might be available in the framework
of the re--acceleration model.

\item[(d)]
The fraction of clusters in our sample with a central radio halo is 
$f_{\rm RH}= 29\pm9$\%. This fraction seems to show some dependence on 
the X--ray luminosity of the hosting cluster. If we define a ``low'' and a 
``high'' X--ray luminosity bin, with a threshold chosen at 
L$_{\rm X} \approx 8\times10^{44}$ erg s$^{-1}$
(for consistency with Cassano et al. \cite{cassano08})
we find that the fraction of clusters with a radio halo 
in the high X--ray luminosity range is as high as 
$f_{\rm RH}=38\pm13$\%.
This result extends to intermediate redshift the findings of Giovannini
et al. (\cite{giovannini99}) for the local Universe (z$\le$ 0.2). 

\end{itemize}

The results of the GMRT radio halo survey have provide observational
support to the re--acceleration scenario for the formation of radio halos.
In such scenario, particles are re--accelerated by 
magneto--hydrodynamic turbulence injected in the ICM during cluster--cluster 
mergers. A basic expectation of this model is that only massive clusters 
undergoing sufficient energetic merger events may host a giant radio halo.
In addition, the finite dissipation time--scale of turbulence implies that 
radio halos should be ``transient'' phenomena, with the duration of the
order of $\sim$ 1 Gyr or less. This is a major difference between
the primary and secondary electron model scenario, the latter predicting that 
radio halos should be very common in galaxy clusters, should be long living
phenomena and are not expected to show any correlation with the cluster
dynamical status.

In this respect the newest and most relevant result of our survey
concerns the undetections and the well constrained upper limits for a large
fraction of the sample objects, at a level which is far below at least
one order of magnitude the radio power expected on the basis of the well
known P$_{\rm 1.4~GHz}$--L$_X$ correlation for giant radio halos in the 
same X--ray luminosity interval. 
This gap indicates a bimodal behaviour in the radio emission properties of 
massive clusters and strongly supports the re--acceleration model whereby 
the giant radio halos are temporarily,
although impressive, events tied up with the recent occurrence of large
mergers, as we have recently discussed (BVD07).

As a final point we would like to stress the importance of our GMRT survey,
which provides new statistical basis on the occurrence  of giant radio
halos in the critical redshift range 0.2--0.4. A combination of our results
with those of the lower redshift samples shows that there is good
statistical evidence (at 3.6$\sigma$) of an increase of the occurrence of giant
radio halos with increasing X-ray luminosity (mass), in agreement with the
prediction of the reacceleration model, as extensively discussed in
Cassano et al. (\cite{cassano08}).
\\
\\
\begin{acknowledgements}
We thank the staff of the GMRT for their help during the observations.
GMRT is run by the National Centre for
Radio Astrophysics of the Tata Institute of Fundamental Research.
We acknowledge financial contribution from the Italian Ministry
of Foreign Affairs, from grants MIUR PRIN2004, PRIN2005 and 2006, 
from PRIN--INAF2005, and from contract ASI--INAF I/023/05/0.
\end{acknowledgements}


\end{document}